\documentclass[twocolumn,showpacs,preprintnumbers,amsmath,amssymb]{revtex4}

\usepackage{graphicx}
\usepackage{dcolumn}
\usepackage{bm}

\begin{document}

\title{Simplified modeling of cluster-shell competition in $^{20}$Ne
and $^{24}$Mg}%

\author{N. Itagaki$^1$, J. Cseh$^2$, and M. P{\l}oszajczak$^3$}
 \affiliation{
$^1$Yukawa Institute for Theoretical Physics, Kyoto University,
Kitashirakawa Oiwake-Cho, 606-8502 Kyoto, Japan\\
$^2$Institute of Nuclear Research of the Hungarian Academy of Sciences, Debrecen, Pf. 51, Hungary-4001\\
$^3$Grand Acc\'{e}l\'{e}rateur National d'Ions Lourds (GANIL), CEA/DSM -- CNRS/IN2P3, BP 5027, F-14076 Caen Cedex 05, France}

\date{\today}

\begin{abstract}
We investigate properties of the Generator Coordinate Method (GCM) on a collective basis of Antisymmetrized Quasi-Cluster (AQC) states to describe the cluster-shell competition in 
$^{20}$Ne and  $^{24}$Mg due to the spin-orbit interaction. By introducing a single additional parameter in the antisymmetrized-cluster basis function, a continuous transformation of the $\alpha$ cluster(s) into independent nucleons can be described. We apply this GCM trial wave function to study in details the transition from cluster states $[ ^{16}$O+$\alpha ]$ and $[ ^{16}$O+$\alpha + \alpha]$ to shell model (SM) states $[ ^{16}$O+${\rm 4N} ]$ and $[ ^{16}$O+${\rm 8N} ]$, respectively. An optimal value of the strength of the spin-orbit interaction is deduced by reproducing level spacing in $^{20}$Ne. A possible connection to the group theoretical understanding of the cluster-shell configuration transition is also discussed.

\end{abstract}

\maketitle

\section{Introduction}

Atomic nuclei provide a rich diversity of properties typical for mesoscopic systems.
One of standard features of these systems is the mean field and the associated shell structure. Indeed, the single-particle motion is a cornerstone of nuclear structure \cite{Boh75} and the stability of nuclei depends on non-uniformities of the single-particle level distribution and presence of magic gaps.  Contrary to atomic systems, a strong spin-orbit interaction is a key ingredient  to fully explain the observed magic numbers in atomic nuclei \cite{Jensen,Mayer}. Another important aspect of nuclear structure is the clustering phenomenon which is of particular importance in states close to their cluster decay thresholds \cite{Ike68} due to generic features of the continuum coupling \cite{Oko03,Dob07}. In this context $\alpha$-particle, which is strongly bound and $\alpha$-$\alpha$ interaction is not strong enough to make a bound state, can be considered as an effective building block of the nuclear structure. Indeed, many features of nuclear masses and excitations in even-even $N=Z$ light nuclei can be explained by (2n-2p) quarteting effects \cite{Danos}. This molecular viewpoint has been introduced by Wheeler, 
von Weizs\H{a}cker and Wefelmeier \cite{Wheeler} long before the nuclear Shell Model (SM) has been proposed \cite{Jensen,Mayer}, and cluster structures have been extensively studied for more than four decades \cite{Hiu63,Tam65,Brink,Supple,Oko,Bau,FMD} (for a selection of recent works, see Ref. \cite{CC}). Recently, theoretical and experimental investigations of clustering phenomena have been extended towards neutron-rich nuclei \cite{Oertzen,Freer} (see Ref. \cite{vOe06} for a recent review).

If an $\alpha$-cluster is expressed as the lowest $(s_{1/2})^4$ SM configuration, it
is a spin-zero system where only the central interactions contributes. The stability of such
a cluster and, hence, its physical significance is closely related to the respective role of
different components of the nuclear interaction in building shell structure and nuclear excitations. Hence, the cluster-shell competition is a key issue to understand how effective building blocks of nuclear structure arise from complex nuclear forces. Usually, this competition is studied at the level of two-nucleon forces only but it is expected that three-nucleon forces and four-nucleon forces may play some role in this problem as well. Another important issue in this discussion is the role of coupling to the decay channels which would require the open quantum system formulation of the many-body problem, as provided by the Continuum Shell Model (for recent reviews see Refs. \cite{Oko03,Vol06,Mic09}). These two important issues will not be considered in this paper. 

Being interested in the shell or cluster nature of a nuclear state
the shell and cluster models, and their interrelation  can guide us.
In general both of these models provide us with a complete (or overcomplete) 
set of basis states, until we speak about the general shell and cluster models,
without severe truncations which are necessarily applied in practical applications
(see e.g. the specific models applied later on in this article).
It means that any state can be expanded in both basis.
Then there are four different possibilities.
(i) The expansion in the shell model basis is simple, but in the cluster basis is complicated;
we can call these states as shell-model-like (SM) states.
(ii) The expansion in the cluster model basis is simple, but in the shell basis is complicated;
we can call these states as cluster-model-like (CM) states
(sometimes they are called rigid molecule-like cluster states).
(iii) The expansion is simple both in the shell model basis and in the cluster basis;
we can call these states as shell-model-like cluster (SMC) states.
(iv) The expansion is complicated both in the shell model basis and in the cluster basis;
these states are not interesting from our present viewpoint.
(For a recent discussion, see Ref. \cite{coc8} and references cited therein). 
In favorable cases, states (i)-(iii) can be characterized by simple symmetries. 
For example, SM states in the superconducting limit obey SU(2) symmetry, 
rigid molecule-like cluster states can have SO(4) symmetry, 
while states with SU(3) symmetry have good shell and cluster nature at the same time.

In this work, we aim at obtaining a transparent picture of the cluster-shell competition based on studies of both binding energies and complex spectra. Hence, even though different microscopic studies of the cluster-shell competition are currently possible \cite{CS,SSS}, we apply here the GCM \cite{Gri57} on a simple, flexible basis of Antisymmetrized Quasi-Cluster (AQC) states \cite{Simple,Masui}, where the cluster-shell transition is described by a single parameter. The AQC parameterization is used here to extract the information about the cluster-shell competition in 
$^{20}$Ne and $^{24}$Mg in various spatial configurations and for different values of the strength of the spin-orbit interaction. 

Recently, a considerable interest is devoted to studying quantum phases and phase-transitions in algebraic models. (For a recent review, see \cite{phase-rev,varna9} and references cited therein.). The transition from a cluster state to the SM state when changing a control-parameter, seems to be analogous to this problematic. Therefore,  we discuss in this paper similarities and differences between the present GCM+AQC approach  and the algebraic approach.  As a result of this comparison, a possible algebraic calculation is outlined  for a further study of the cluster-shell competition.

This paper is organized as follows. The formulation of the model and details of the AQC approach are summarized in section II. In section III, we present results concerning the cluster-shell competition in $^{20}$Ne and $^{24}$Mg. In particular, we exhibit the evolution of  wave functions for different low spin states of $^{20}$Ne with the strength of the spin-orbit coupling by plotting the squared overlap between the variational GCM wave function and the AQC basis states.
Possible connections to a group theoretical understanding of the cluster-shell transition is discussed in section IV. Finally, the main conclusions of this work are given in section V.

\section{The model}
In this section, we introduce the GCM+AQC approach and the many-body Hamiltonian used in the cluster-shell transition studies of this work. Let us begin by presenting the basic idea of the AQC wave function and its relation to other parameterized cluster-like wave functions.

\subsection{The Brink-Bloch wave function}
In conventional $\alpha$-cluster models, the single particle wave function is described as
a Gaussian packet \cite{Brink}:
\begin{equation}
\psi_i = \left( {2\nu \over \pi}\right)^{3 \over 4}
\exp[-\nu(\vec r_i-\vec R_{\gamma}/\sqrt{\nu})^2] \chi_i ~ \ ,
\end{equation}
where $\chi_i$ represent the spin-isospin part of the 
$i$-th single particle,
and $R_{\gamma}$ is a real parameter representing the center of a Gaussian for  
$\gamma$-th $\alpha$-cluster. 
To assure that  the spurious center-of-mass kinetic energy is exactly removed, the parameter $\nu$ is the same for {\em all} nucleons.
In this Brink-Bloch wave function \cite{Brink}, four nucleons (spin-up proton/neutron, spin-down proton/neutron) share the common $R_{\gamma}$ value. Hence, the spin-orbit interaction vanishes for Brink-Bloch $\alpha$-clusters. In general, even if there are valence nucleons around clusters, the spin-orbit interaction vanishes for a single Slater determinant wave function because the time-odd components in the spatial part of such a wave function are missing \cite{explanation}. Hence, the contribution of 
the spin-orbit interaction for valence nucleons around 
$\alpha$-cluster(s) can be taken into account only by performing the angular momentum projection and/or by superposing the Slater determinants with complex coefficients. However, for 
$(N\alpha)$-nuclei, the spin-orbit interaction does not act even after these treatments, since all  clusters have spin zero.

\subsection{The AMD and FMD wave functions}
In AMD \cite{Ono1} and Fermionic Molecular Dynamics (FMD) \cite{FMD1} approaches, both deriving from the Time-Dependent Cluster (TDC) approach \cite{Drozdz}, the Gaussian center parameters are allowed to be complex. In this case, the single particle wave functions contain time-odd components, and the spin-orbit interaction acts even at the level of a single Slater determinant, i.e. before the angular momentum projection. The single particle wave function of 
$i$-th nucleon has the following form:
\begin{equation}
\psi_i = \left( {2\nu \over \pi}\right)^{3 \over 4}
\exp[-\nu(\vec r_i-\vec z_i/\sqrt{\nu})^2] \chi_i ~ \ ,
\label{amd}
\end{equation}
where $\vec z_i$ is a complex parameter \cite{explanation1}. The real and imaginary parts of 
$\vec z_i$ represent the expectation values of the position: $<\vec r> =  {\rm Re}[ \vec z_i ] / \sqrt{\nu}$, and momentum: $<\vec p> = 2 \sqrt{\nu} \hbar \ {\rm Im}[\vec z_i]$, of the $i$-th nucleon. Since this single particle wave function breaks the time-reversal symmetry, it allows to describe the boost of nuclei. Thus, this variational ansatz (\ref{amd}) for the wave function can be applied not only in static, nuclear structure calculations but also in dynamical nuclear reaction calculations \cite{Oko,Bau,Ono1,FMD2}. In the latter case, equations of motion for the time-evolution of 
parameters $\{\vec z_i\}$ \cite{Drozdz} are derived from the variational principle associated with the time-dependent Schr\"{o}dinger equation \cite{Koonin}.

\subsection{The AQC wave function}
The AQC approach \cite{Simple,Masui} allows for a convenient transformation of the SM wave function in $jj$ coupling scheme into the $\alpha$-cluster wave function. Hence, this approach is particularly well suited for studies of the cluster-shell competition and a description of the mixed phase in terms of quasi-clusters, i.e. the SM-like cluster states.
Let us suppose that the nucleus consists of few $\alpha$-clusters and one quasi-cluster. In general, one can also consider valence nucleons but to keep the discussion of the cluster-shell competition as transparent as possible, we shall restrict ourselves to systems consisting of $\alpha$-clusters and quasi-cluster(s).
The Gaussian-center parameters $\vec z_i/\sqrt{\nu}$ for nucleons in $\alpha$-clusters are real numbers. For nucleons in the quasi-cluster, the single-particle wave function is described as a
Gaussian packet:
\begin{equation}
\psi_i = \left( {2\nu \over \pi}\right)^{3 \over 4}
\exp[-\nu(\vec r-\vec \zeta_i/\sqrt{\nu})^2] \chi_i ~ \ ,
\label{GPG}
\end{equation}
where ${\vec \zeta_i}/\sqrt{\nu}$ are the generalized center parameters of the packet, and:
\begin{equation}
\vec \zeta_i  =  \vec z_i + i \Lambda (\vec e^{i}_{spin}) \times {\rm Re}[\vec z_i] ~ \ .
\label{cenpar}
\end{equation}
In the above equation, $\vec e^i_{spin}$ is the unit vector for the intrinsic-spin orientation, and 
$\Lambda$ is a control parameter describing the dissolution of the (quasi)-cluster. 

The spin-orbit interaction is intuitively interpreted as $(\vec r \times \vec p) \cdot \vec s$ and this is equal to $(\vec s \times \vec r) \cdot \vec p$, where $\vec r$, $\vec p$, and $\vec s$ represent the position, the momentum, and the spin of the nucleon, respectively. If nucleons of the quasi-cluster have the momentum components parallel (anti-parallel) to $\vec s \times \vec r$, the spin-orbit interaction acts attractively (repulsively), i.e.  the spin-orbit coupling acts on each individual wave function. For  positive (negative) $\Lambda$, the contribution of spin-orbit interaction is attractive (repulsive). In actual calculation, of course, the spin-orbit coupling is a two-body force, so this interaction is a function of ${\vec r}_i-{\vec r}_j$ and not of ${\vec r}$.

Parameters $\{\vec z_i\}$ and/or $\Lambda$ are complex in the time-dependent formulation of nuclear reactions and dynamics. In the present studies, however, we perform static calculations, so $\{ \vec z_i\}$ and $\Lambda$ are real. Hence, real and imaginary parts of the Gaussian-center parameters for nucleons of the quasi cluster are:
\begin{eqnarray}
{\rm Re}[\vec \zeta_i] &=& \vec z_i \nonumber \\
{\rm Im}[\vec \zeta_i] &=& \Lambda (\vec e^{i}_{spin}) \times \vec z_i ~ \ . \nonumber
\end{eqnarray}
The real parts of $\{ \vec \zeta_i \}$, are the same for four nucleons of the quasi cluster. 

In the following sections, this parameterized wave function will be applied to explain observed levels in $^{20}$Ne and an evolution of the wave function with the strength of the spin-orbit coupling in 
$^{20}$Ne and $^{24}$Mg. In particular, it will be shown that the limits $\Lambda = 0$ and 
$\Lambda = 1$ of the AQC wave function correspond to the cluster wave function and the SM wave function (the spherical harmonics), respectively. $\Lambda < 0$ describes a particle in spin-orbit unfavored orbits, whereas $\Lambda > 1$ corresponds to a particle in higher shells. 

\subsection{The Hamiltonian}
The Hamiltonian operator $(\hat{H})$ has the following form:
\begin{equation}
\hat{H}=\sum_{i=1}^{A}\hat{t}_i-\hat{T}_{c.m.}+\sum_{i>j}^{A}\hat{v}_{ij},
\end{equation}
where a two-body interaction $(\hat{v}_{ij})$ includes the central part, the spin-orbit part and the Coulomb part. For the central part, we use the Volkov2 effective $N-N$ potential \cite{Vol}:
\begin{eqnarray}
V(r)=&& (W - MP^\sigma P^\tau) \nonumber \\
     && \times \left[V_1\exp(-r^2/c_1^2)+V_2\exp(-r^2/c_2^2)\right] ,
\end{eqnarray}
where $W=1-M$, $V_1^c = -60.65$ MeV, $V_2^c = 61.14$ MeV, $c_1 = 1.8$ fm, and 
$c_2 = 1.01$ fm. For the spin-orbit term, we introduce the two-range variant of the G3RS potential \cite{G3RS}:
\begin{equation}
V_{ls} = V_0 (e^{-d_1 r^2} - e^{-d_2 r^2} ) P(^3{\rm O}){\vec L}
\cdot {\vec S},
\end{equation}
where $d_1=5.0$ fm$^{-2}$, $d_2=2.778$ fm$^{-2}$, and $P(^3{\rm O})$ is a projection operator onto a triplet-odd state. The operator $\vec L$ stands for the relative angular momentum
and $\vec S\equiv\vec{S_1}+\vec{S_2}$ is the spin operator.

The parameter set: $M = 0.60$ and $V_0 = 2000$ MeV, is known to give a reasonable description of $\alpha+n$ and $\alpha+\alpha$ scattering phase shifts \cite{Okabe79}. For heavier nuclei, beyond $^{12}$C, one needs however larger $M$ values in the structure calculations. 
In the present study of energy levels in $^{20}$Ne and $^{24}$Mg, we take $M = 0.62$ and compare experimental and calculated spectra for three values of the strength parameter $V_0 = 0, 1000, 2000$ MeV.

\subsection{The GCM trial wave function}
The trial wave function of an A-particle system is constructed in the form:
\begin{eqnarray}
&& \Phi({\vec r}_1,\cdots,{\vec r}_A) = \nonumber \\
&& \int\Psi({\vec r}_1,\cdots,{\vec r}_A;\Lambda, R)f(\Lambda,R) \ d\Lambda dR
\label{eq1}
\end{eqnarray}
where $f(\Lambda,R)$ is a variational function of GCM and 
$\Psi({\vec r}_1,\cdots,{\vec r}_A;\Lambda, R)$ is a basis function:
\begin{eqnarray}
&& \Psi({\vec r}_1,\cdots,{\vec r}_A;\Lambda,R) = \nonumber \\
&& {\hat P}^\pi {\hat P}^J_{MK} \ {\cal A}
[ (\psi_1(\vec r_1)\chi_1) \cdots (\psi_k(\vec r_k)\chi_k) \\
&&   \ \ \ \ (\psi_{k+1}(\vec r_{k+1};\Lambda,R)\chi_{k+1}) 
\cdots (\psi_A(\vec r_A;\Lambda,R)\chi_A)].
\end{eqnarray}
${\hat P}^\pi$ and ${\hat P}^J_{MK}$ in (\ref{eq1}) are projection operators on a good parity and angular momentum, respectively, and
$\psi_i$ and $\chi_i$ represent the spatial and spin-isospin parts of the 
$i$-th single particle AQC wave function, respectively. 
The single particles from 1 to $k$ belong to normal cluster(s) and
those from $k+1$ to $A$ belong to quasi cluster(s).
Here,
$\Lambda$ is a control parameter introduced for the single particle wave functions
which belong to the quasi-cluster(s), and $R$
is a distance parameter between the quasi cluster and center of mass of other clusters
(which will be introduced later). 

The preliminary A-particle wave function $\Psi({\vec r}_1,\cdots,{\vec r}_A;\Lambda,R)$ depends upon a collective parameters or generator coordinates, $\Lambda$ and $R$. The collective wave function, 
$f(\Lambda,R)$, is folded into $\Psi$ to produce a system wave function that depends only on the nucleonic coordinates ${\vec r}_i$. The collective parameter $\Lambda$ and $R$ generate the trial A-particle wave function but disappears in the final state function $\Phi({\vec r}_1,\cdots,{\vec r}_A)$. For a given A-particle Hamiltonian, the projected GCM trial wave function (\ref{eq1}) is then expressed with respect to the generator function $f(\Lambda,R)$.

\section{Results}
In this section, we investigate the cluster-shell competition due to the spin-orbit interaction using the GCM on a collective basis generated by the AQC basis states. The GCM  studies will be applied in Sect. III.A for low-lying states of $^{20}$Ne, described as $^{16}$O-core and one 
quasi-$\alpha$ cluster, and in Sect. III.B for states of $^{24}$Mg given by $^{16}$O-core and two 
quasi-$\alpha$ clusters.

\subsection{$^{20}$Ne}
Numerous calculations exist for $^{20}$Ne within the model space of $^{16}$O+$\alpha$ or 
$^{12}$C+$\alpha$+$\alpha$ \cite{Matsuse,Kato,Fujiwara}. From the SM point of view, two protons (two neutrons) outside of the $^{16}$O core occupy the $\pi(d_{5/2})$ ($\nu(d_{5/2})$) orbit,  and the spin-orbit interaction acts strongly \cite{Akiyama,McGrory}. This effect cannot be taken into account in a simple cluster-model space. Hence, the hybrid models 
\cite{Tomoda,Suzuki} and the AMD approach \cite{EnyoNe,Taniguchi} have been applied to describe both the cluster structures and the single-particle motion of nucleons around the $^{16}$O-core. Finally, the $^{16}$O-core+$3N+N$ three-cluster approach has been proposed to account for the breaking of one $\alpha$-cluster due to the spin-orbit interaction \cite{Yamada}.

In our model of $^{20}$Ne, four $\alpha$-clusters form a tetrahedron which corresponds to the doubly closed $p$-shell if the relative distance between clusters is equal to zero. The actual distance between $\alpha$-clusters in the $^{16}$O-tetrahedron is 1.0 fm, and its center of mass is placed at the origin of the coordinate system. The remaining (quasi-)$\alpha$ cluster is located on the $x$-axis. The center of mass of this cluster is  denoted by $R\vec e_x$, where $R$ is  a distance parameter and $\vec e_x$ is the unit vector on the $x$-axis.

If the spin-orbit interaction acts, the quasi-$\alpha$ cluster has $S \neq 0$ because the spatial part of the wave function for each nucleon of the quasi-cluster is not an eigenstate of the time-reversal operator.  The dissolution of $\alpha$-cluster ($S=0$ ) by the spin-orbit interaction is described by the control parameter (the generator coordinate) $\Lambda$ in the AQC basis wave function. Since the direction of the spin is defined along the $z$-axis, we change the Gaussian centers of the nucleons in the quasi-cluster from $R \vec e_x$ to $R (\vec e_x + i \Lambda \vec e_y)$ or 
$R (\vec e_x - i \Lambda \vec e_y)$, for spin-up or spin-down nucleons, respectively, to assure that the directions of the spin and orbital parts of the angular momentum are parallel. Hence, the generalized center parameters of the quasi-$\alpha$ packet become:
\begin{equation}
{\vec \zeta}/\sqrt{\nu} = 
R (\vec e_x + i \Lambda \vec e_y) ~ \ ,
\label{sup}
\end{equation}
and 
\begin{equation}
{\vec \zeta}/\sqrt{\nu} = 
R (\vec e_x - i \Lambda \vec e_y) ~ \ ,
\label{sdown}
\end{equation}
for the for the spin-up and spin-down nucleons (proton or neutron), respectively. Here, 
$R \vec e_x$ is the spatial position of the Gaussian center. Imaginary parts: $R \Lambda \vec e_y$ and $-R \Lambda \vec e_y$, express the momenta of nucleons, and $\vec e_x$ and $\vec e_y$ are unit vectors on $x$ and $y$ axes, respectively. 

These imaginary parts of the generalized center parameters of the quasi-$\alpha$ packet allow to mimic spherical harmonics with a parametrization (\ref{GPG}). The spatial part of  the single-particle wave function (\ref{GPG}) of a nucleon in the quasi cluster is:
\begin{eqnarray}
\psi_i = 
\left( {2\nu \over \pi}\right)^{3 \over 4}
\exp[-\nu \vec r^2 -{\vec \zeta}^2 + 2\nu \vec r \cdot \vec 
\zeta /\sqrt{\nu} ] ~ \ . 
\end{eqnarray}
The last term 
can be expanded using (\ref{sup}), (\ref{sdown}). In the case of a spin-up nucleon, for example, one obtains:
\begin{eqnarray}
\exp[2\nu \vec r \cdot \vec \zeta /\sqrt{\nu} ] = 1+\sum_{k=1}^{\infty}\frac{1}{k!}(2 \nu R (x + i \Lambda y))^k ~ \ .
\end{eqnarray}
For $\Lambda = 1$, one finds:
\begin{eqnarray}
\exp[2\nu \vec r \cdot \vec \zeta /\sqrt{\nu}] = 1+\sum_{k=1}^{\infty}\frac{1}{k!}\frac{1}{s_k}(2\nu rR)^k Y_{kk}(\Omega) ~ \ ,
\end{eqnarray}
where $s_k$ are the normalization factors of spherical harmonics $Y_{kk}(\Omega)$.
Hence, the spatial part of the wave function of a spin-up nucleon in the quasi cluster can be written as: 
\begin{eqnarray}
\psi_i =  \left( {2\nu \over \pi}\right)^{3 \over 4}
\{1&+&s_1^{-1} 2\nu  r_iR Y_{11}(\Omega_i) \nonumber \\
  &+& (1/2!) s_2^{-1} (2\nu r_iR)^2 Y_{22}(\Omega_i) \nonumber \\ 
  &+& (1/3!) s_3^{-1} (2\nu r_iR)^3 Y_{33}(\Omega_i) \nonumber \\
  &+&\cdots + (1/n!) s_n^{-1} (2\nu r_iR)^n Y_{nn}(\Omega_i) \nonumber \\
  &+&\cdots \} \exp[-\nu r_i^2]. 
\end{eqnarray}
Since the direction of spin is along the $z$-axis,
the spin-up wave function is described as a linear combination of 
$|j\ j_z=+j\rangle$ states: 
\begin{eqnarray}
a_{1/2,1/2}|1/2\ 1/2\rangle &+& a_{3/2,3/2}|3/2\ 3/2 \rangle \nonumber \\ &+& a_{5/2,5/2}|5/2\ 5/2 \rangle 
+ \cdots ~ \ .
\nonumber
\end{eqnarray}
Analogously, the spatial part of the wave function for a spin-down nucleon is:
\begin{eqnarray}
a_{1/2,-1/2}|1/2\ -1/2\rangle &+& a_{3/2,-3/2}|3/2\ -3/2 \rangle \nonumber \\ &+& a_{5/2,-5/2}|5/2\ -5/2 \rangle 
+ \cdots ~ \ .
\nonumber
\end{eqnarray}
This can be obtained by a time-reversal transformation of the spin-up wave function.

\begin{figure}
\includegraphics[width=7.5cm]{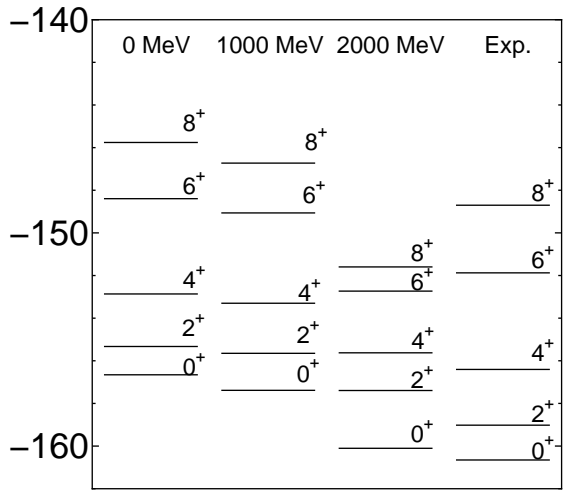} 
\caption{
The GCM calculations of yrast levels in $^{20}$Ne for three different values of the strength of the spin-orbit interaction: $V_0 = 0, 1000, 2000$ MeV, are compared with the experimental
data (Exp.). For more details, see the description in the text.}
\end{figure}
\begin{figure}
\includegraphics[width=6.5cm]{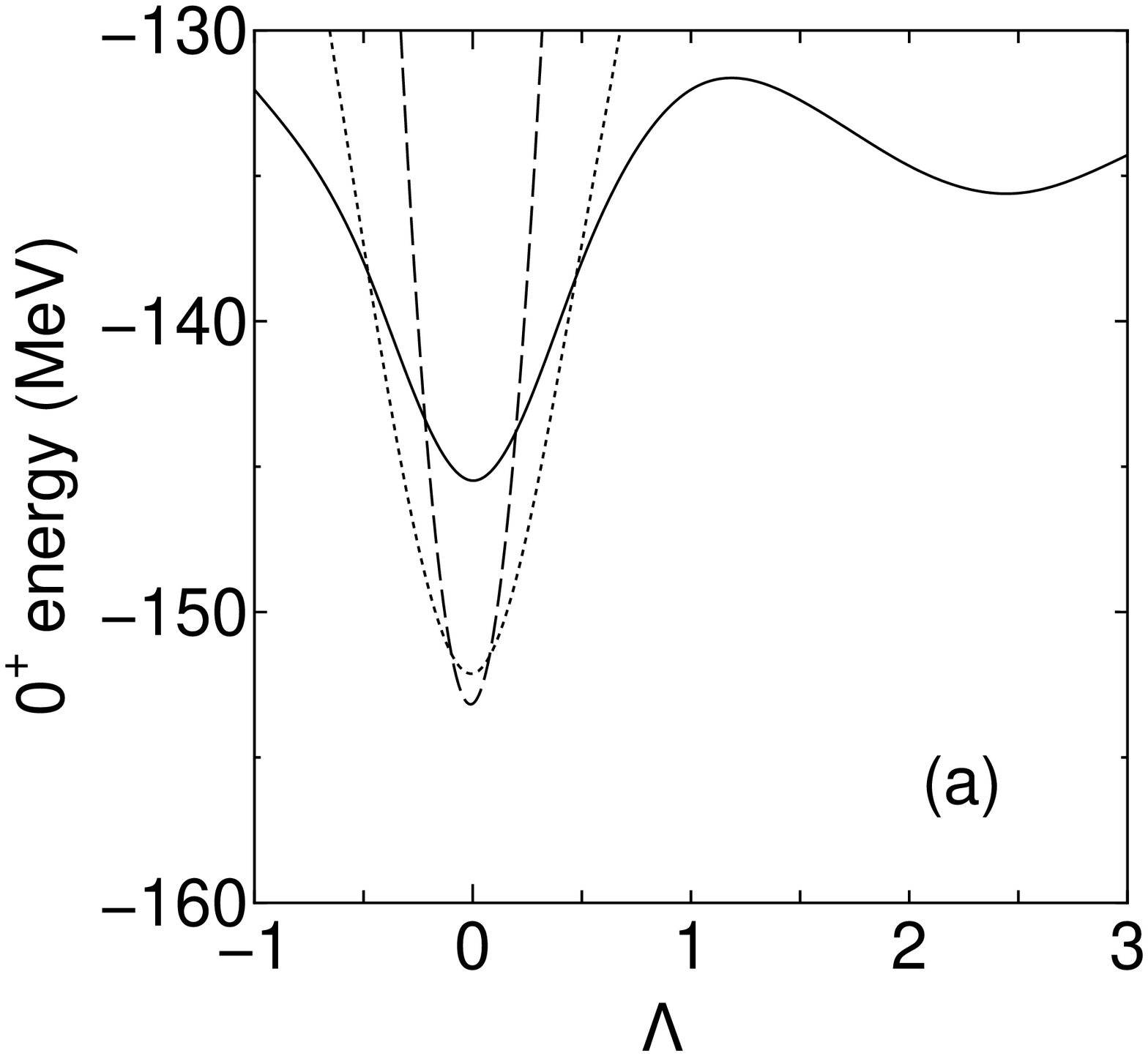} 
\includegraphics[width=6.5cm]{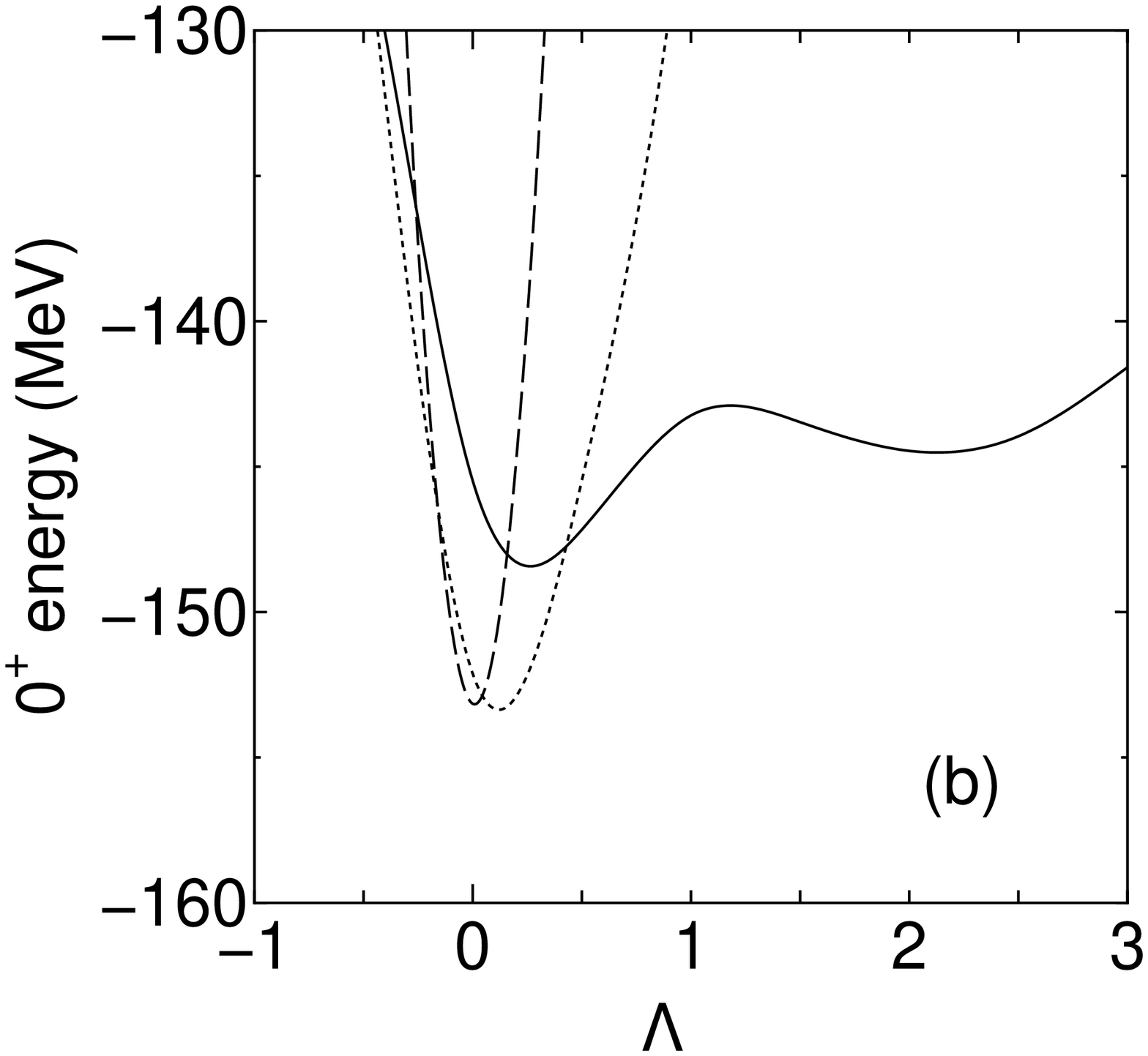}
\includegraphics[width=6.5cm]{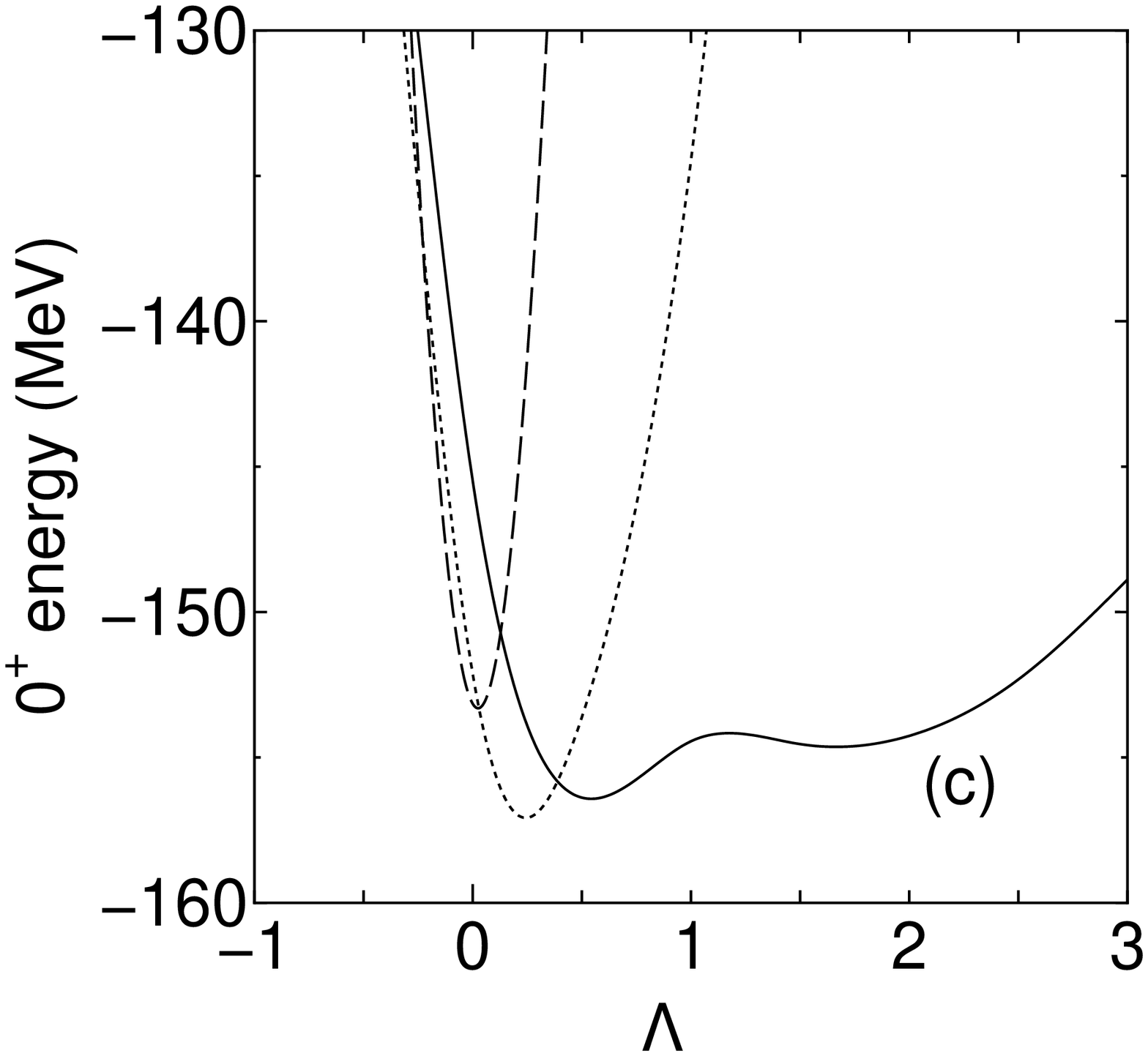}  
\caption{
The energy curves of the $0^+$ state
of $^{20}$Ne as a function
of $\Lambda$: 
(a): $V_0$ = 0 MeV,
(b): $V_0$ = 1000 MeV, and
(c): $V_0$ = 2000 MeV.
The solid, dotted, 
and dashed lines show
$R = $ 0.5, 2, and 4 fm, respectively.
}
\end{figure}
\begin{figure}
\includegraphics[width=6.5cm]{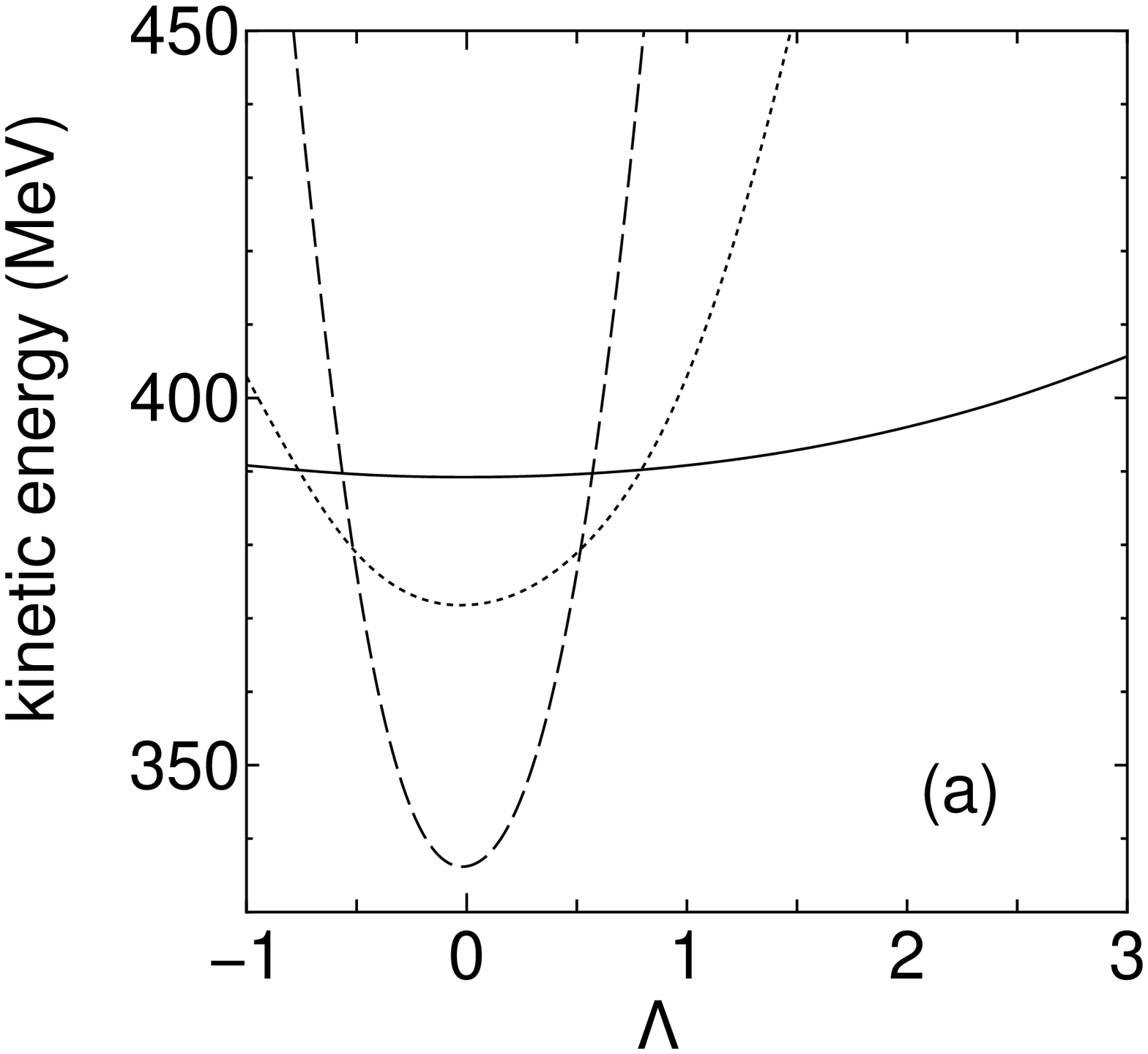} 
\includegraphics[width=6.5cm]{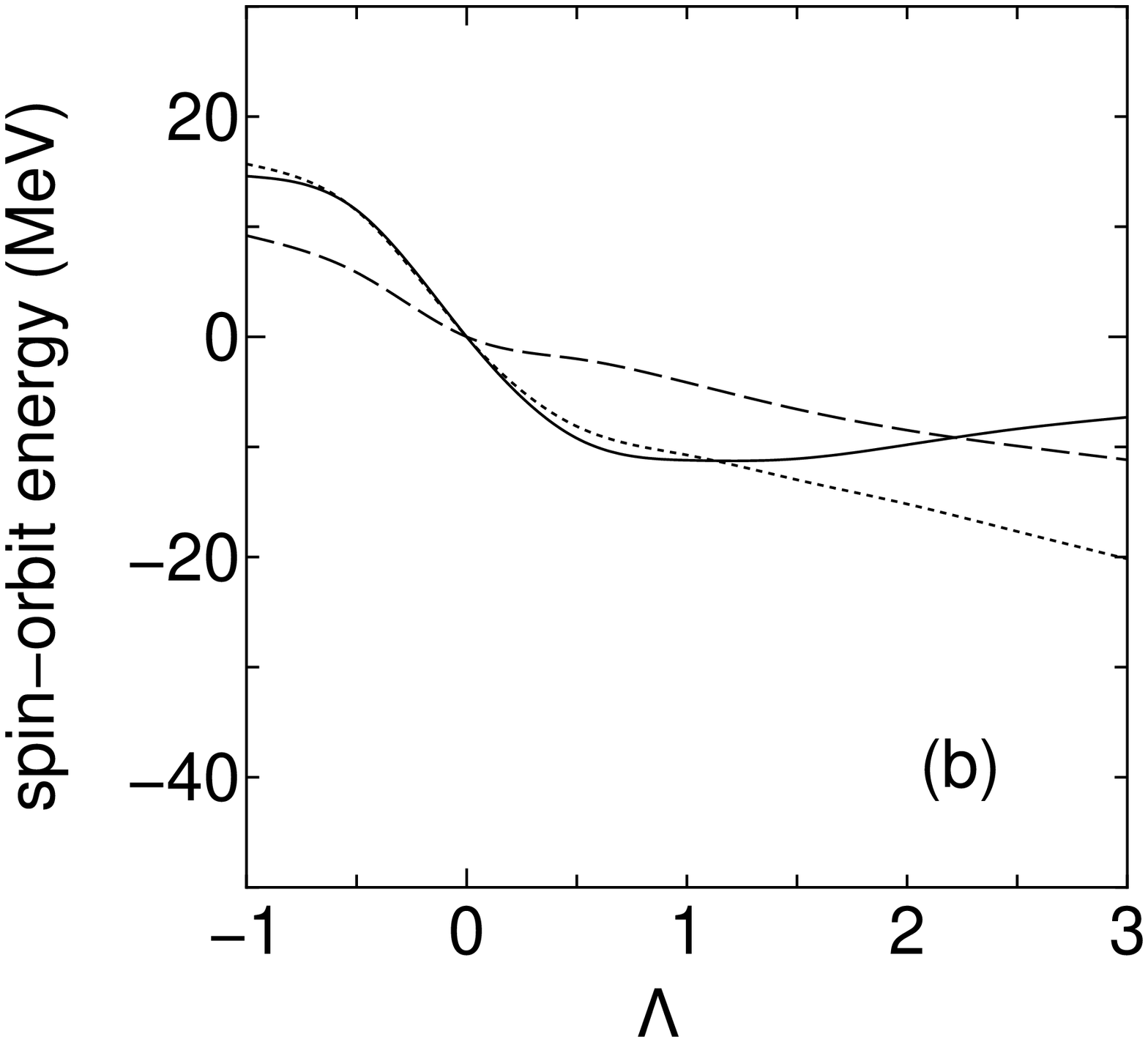}
\includegraphics[width=6.5cm]{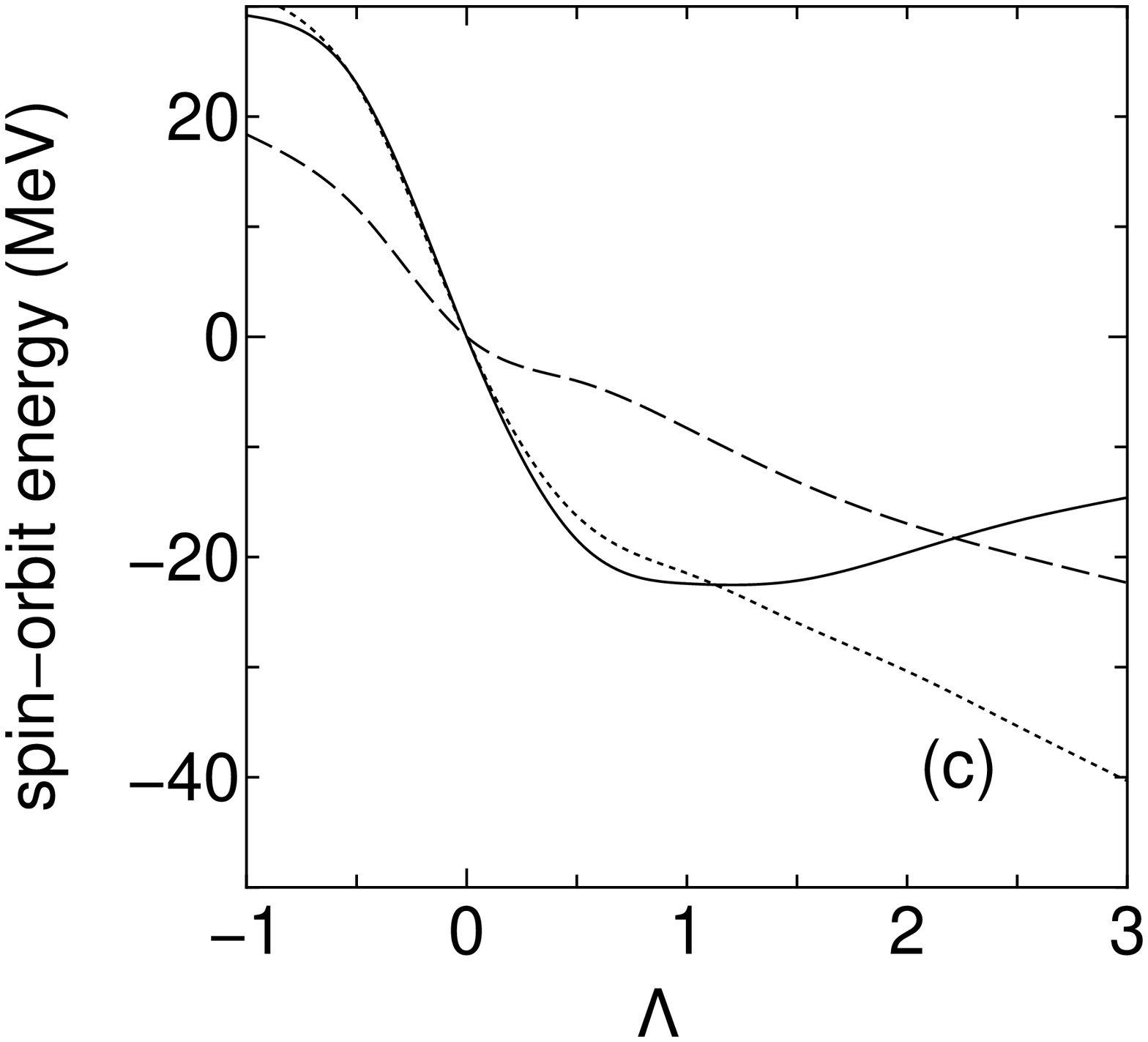}  
\caption{
The kinetic and spin-orbit energy curves of the $0^+$ state
of $^{20}$Ne as a function
of $\Lambda$: 
(a): kinetic energy (independent of the spin-orbit strength),
(b): spin-orbit energy for $V_0$ = 1000 MeV, and
(c): spin-orbit energy $V_0$ = 2000 MeV.
The lines are the same as in Fig. 2.}
\end{figure}

\begin{figure}
\includegraphics[width=6.5cm]{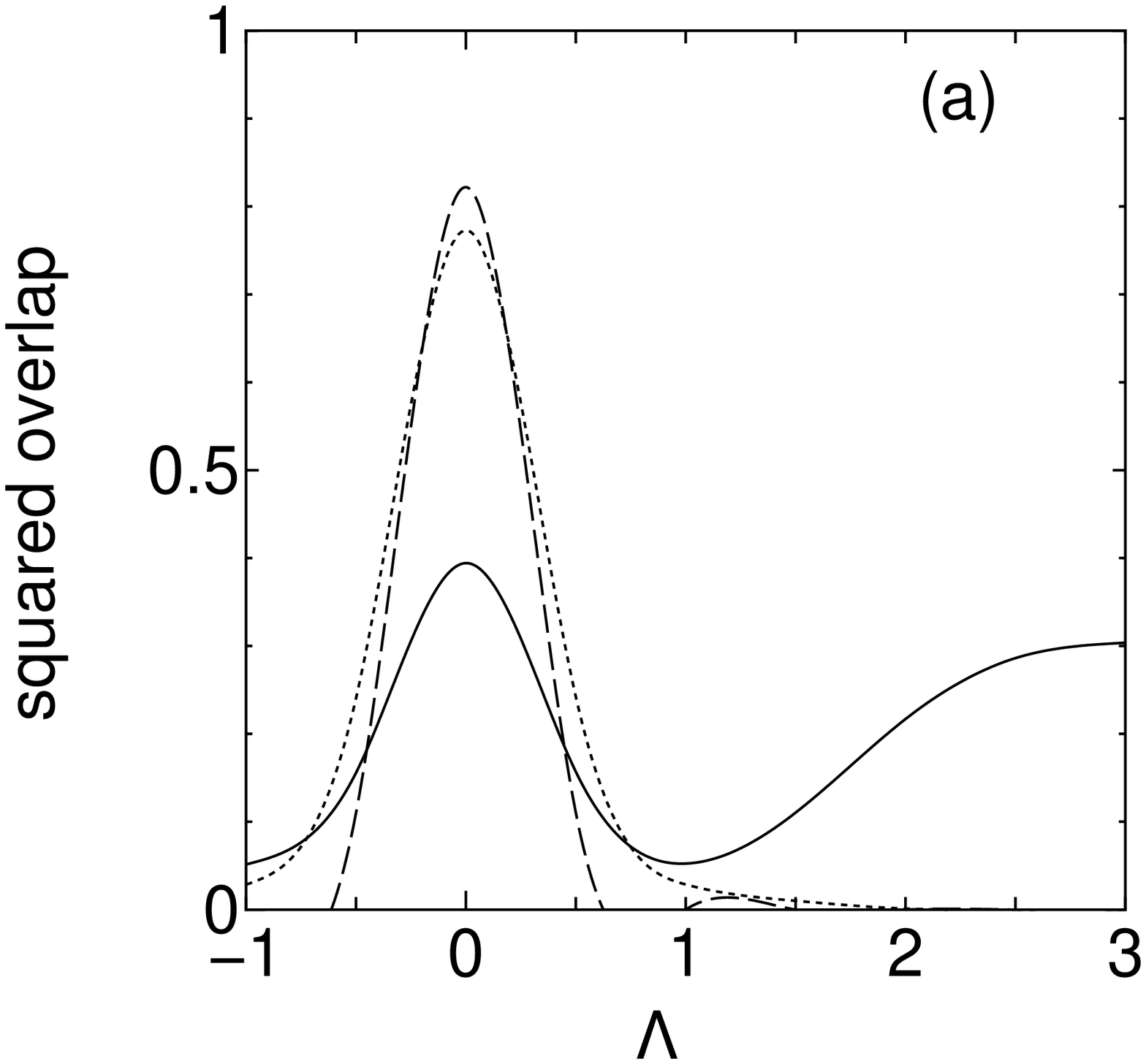} 
\includegraphics[width=6.5cm]{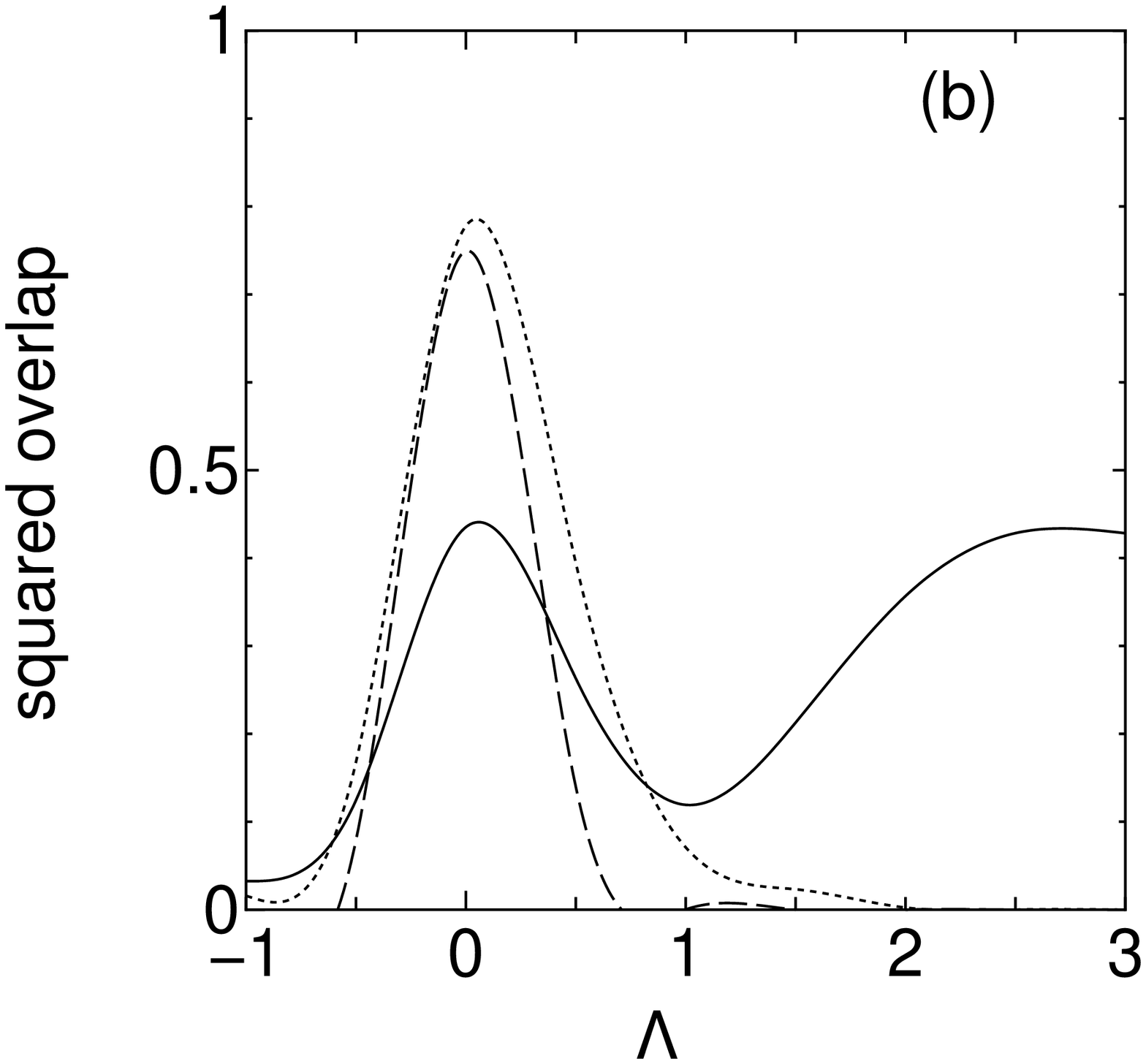}
\includegraphics[width=6.5cm]{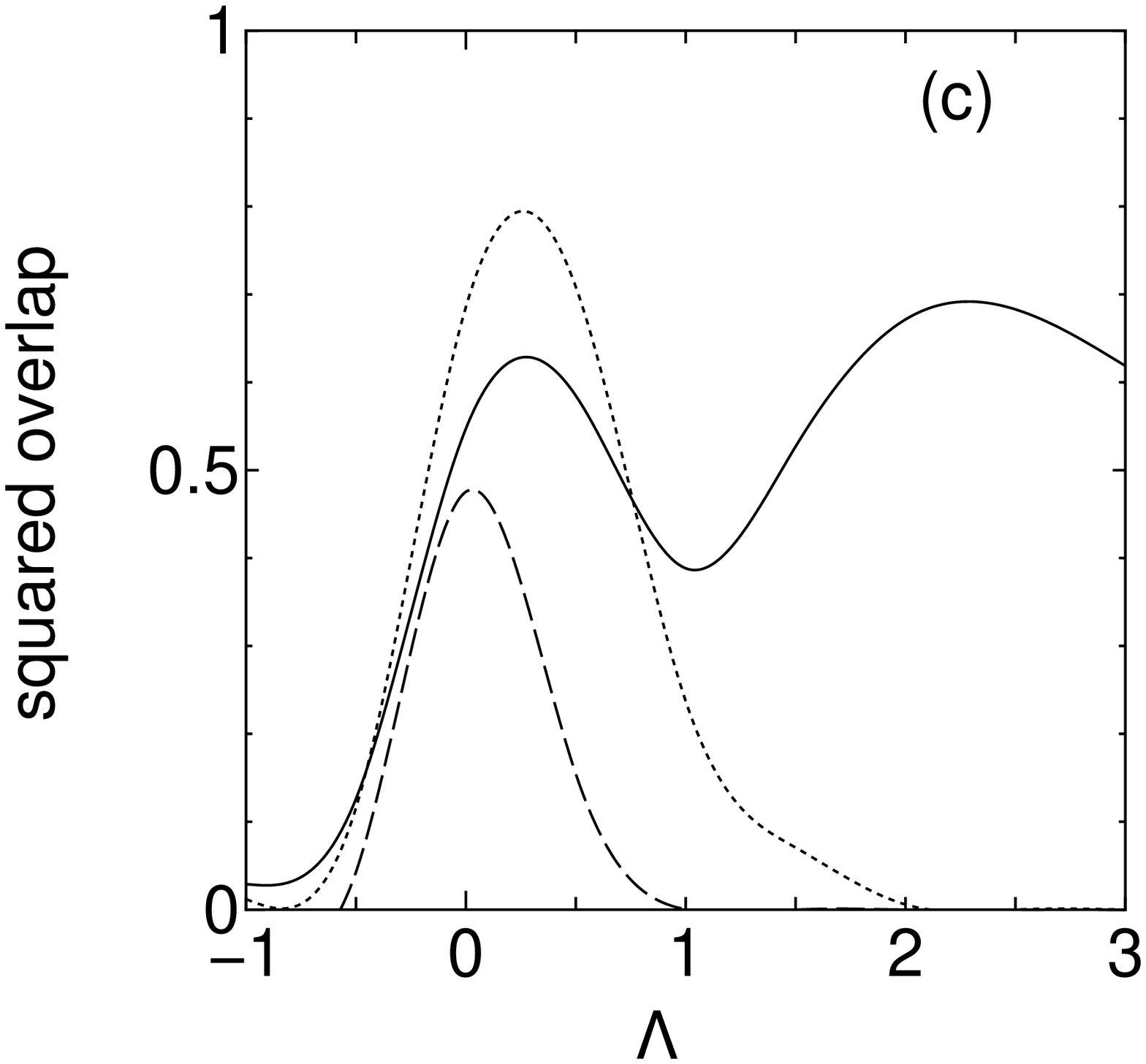}  
\caption{
The squared overlap between the lowest
$0^+$ state of $^{20}$Ne
obtained by superposing the basis states
with different $R$ and $\Lambda$ values
and each basis state.
(a): $V_0$ = 0 MeV,
(b): $V_0$ = 1000 MeV, and
(c): $V_0$ = 2000 MeV.
The lines are the same as in Fig. 2.}
\end{figure}
\begin{figure}
\includegraphics[width=7.5cm]{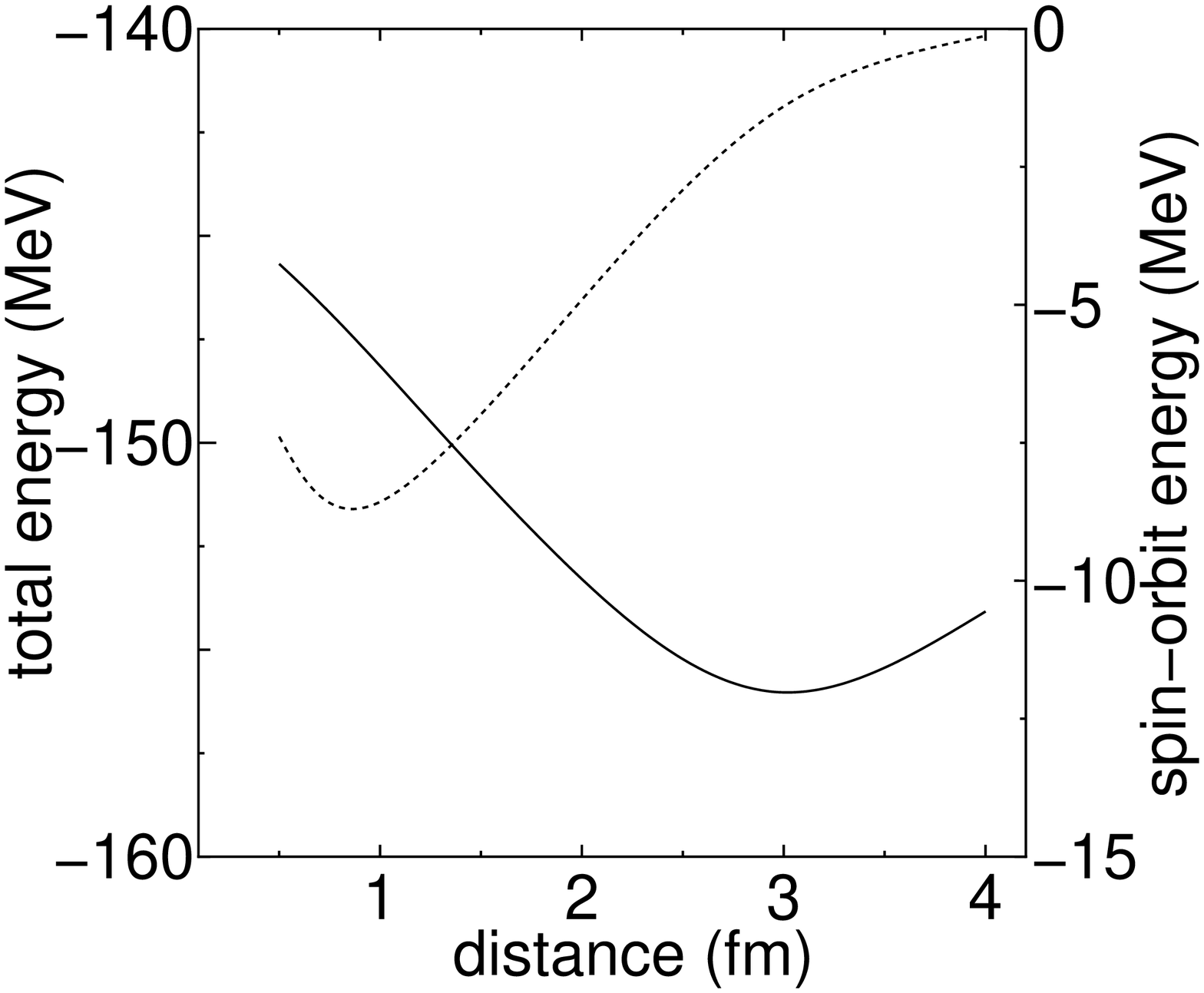}    
\caption{
The energy curves of the $0^+$ state
as a function of the distance between ``quasi cluster"
and $^{16}$O is shown.
The solid line is total energy (left vertical axis), 
and the dotted line is spin-orbit energy (right vertical axis).
The strength of the spin-orbit interaction
is $V_0$ = 1000 MeV, which gives reasonable level
spacing in Fig. 1.}
\end{figure}
It is important to mention that all components of the quasi cluster wave function
which overlap with the wave function of the core nucleus are exactly removed by the Pauli principle.
In the case of $^{20}$Ne, four nucleons of the quasi cluster are excited to the $sd$-shell. For small $R$, the radial wave function for the spin-up proton/neutron corresponds to the $r^2Y_{22}\exp[-\nu r^2]$ SM wave function, whereas for the spin-down proton/neutron one finds $r^2Y_{2-2}\exp[-\nu r^2]$ SM radial wave function. Therefore, $\Lambda$ is a parameter which can be used to characterize the
cluster-shell competition: $\Lambda$ = 1 is a limit of the spherical harmonics, and $\Lambda$ = 0 is the $\alpha$-cluster limit. If $R$ is large, the radial wave function contains admixtures of the higher shell components.

After setting Gaussian center parameters of nucleons in quasi cluster and $\alpha$-clusters,
the whole system is Galilei transformed so that the center of gravity coincides with the origin of 
the coordinate system.

Fig. 1 shows energies of yrast states in $^{20}$Ne. We superpose basis states with different values of 
$\Lambda$ ($\Lambda =$ $-1$, $-0.5$, 0, 0.5, 1.0, 1.5, 2, 2.5, 3)
and $R$ ($R =$ 0.5, 1, 2, 3, 4 fm) generator coordinates, and then diagonalize the Hamiltonian in this subspace. 
In this calculation, the Majorana parameter is 
$M = 0.62$ and the size of the Gaussian wave-packet equals 
$b = 1.6$ fm $(b \equiv { 1 / \sqrt{2\nu}})$. 
For this choice of the parameters, one can reproduce the binding energy of $^{20}$Ne reasonable well.
The results with the strength of the spin-orbit interaction
of $V_0 = 0, 1000,$ and 2000 MeV are compared with experimental
one (Exp.).
The experimental level spacing 
among $0^+$, $2^+$, $4^+$ shows the deviation 
from the $l(l+1)$ rule for the rigid rotor, 
and the spin-orbit interaction is needed to explain
this deviation; 
the experimental $0^+-2^+$ level spacing
is larger than the results with $V_0 = 0$ MeV
even if the $0^+-4^+$ spacing is almost reproduced.
It has been known that introducing 
the spin-orbit strength of $V_0 = 2000$ MeV 
gives reasonable description
for the scattering phase shift of $\alpha$+$N$,
however this strength is too strong for $^{20}$Ne
judging from the obtained level spacing between the ground
$0^+$ state and $8^+$ state.
Therefore, the strength of $V_0 = 1000$ MeV is considered to 
be the best to reproduce
both the $0^+-2^+-4^+$ and $0^+-8^+$ level spacings.
Both the experimental result and the calculation
show that the $8^+$ state does not fit to 
the sequence of the rotational band structure.
This is due to the shell effect, and the deviation
becomes larger with increasing the strength
of the spin-orbit interaction.

One of the possible reasons why $V_0 = 2000$ determined
from the $\alpha$+$N$ scattering is too strong for $^{20}$Ne
is the contribution of the tensor interaction.
It has been known that the tensor interaction 
causes the two-particle-two-hole excitation of $^4$He
from the $(0s)^4$ configuration and this effect plays
an essential role for explaining the 
spin-orbit splitting of $^5$He \cite{Terasawa,Myo}.
Thus we somehow overestimate 
the spin-orbit strength
when we try to fit the scattering phase shift of $^4$He+$N$
using $(0s)^4$ configuration for $^4$He and
without introducing the tensor interaction explicitly.
Also, the first order term of the tensor interaction
(between $j$-upper protons and $j$-upper neutrons)
acts repulsively \cite{Tensor}.
In the present case, 
an $\alpha$-cluster is broken to take into account the spin-orbit
interaction, and two protons and two neutrons occupy 
the $j$-upper orbits, and the first order term of the tensor interaction
acts repulsively between protons and neutrons.
Renormalization of this tensor effect into the spin-orbit part
is another reason for
the reduction of the strength of the spin-orbit interaction.

The energy curves of the $0^+$ state as a function
of $\Lambda$, which are expressed as
\[
\langle
\Psi({\vec r}_1,\cdots,{\vec r}_A;\Lambda, R) 
|\hat{H}|
\Psi({\vec r}_1,\cdots,{\vec r}_A;\Lambda, R)
\rangle 
\ / \
\langle \Psi | \Psi \rangle 
\]
in terms of Eq. (9),
are shown in Fig. 2, where
(a): $V_0$ = 0 MeV,
(b): $V_0$ = 1000 MeV, and
(c): $V_0$ = 2000 MeV.
The lines are the results with different distance
between $^{16}$O and quasi $\alpha$-cluster,
and
the solid, dotted,
and dashed lines are cases of 
$R = $0.5, 2, and 4 fm, respectively.
In the case of $V_0 = 0$ MeV ((a)),
the minimum points of the curves appear 
around $\Lambda = 0$.
This is because
there is no spin-orbit interaction in this case.
Introducing $\Lambda$ value induces
the rotation of four nucleons in the 
``quasi $\alpha$" cluster around $^{16}$O, 
which increases the kinetic energy of the
four nucleons.
Thus, without spin-orbit interaction,
the energy does not decrease.
On the other hand,
with increasing the $V_0$ value,
the minimum points shifts to finite $\Lambda$
value ((b) and (c)).
The lowest energy is obtained with $R = 3$ fm
and $\Lambda \sim$ 0.2 in the case of 
$V_0 = 1000$ MeV 
(although the line for $R = 3$ fm is not shown in this figure).

The kinetic and spin-orbit energy curves of the $0^+$ state
of $^{20}$Ne as a function
of $\Lambda$ are shown in Fig. 3, where
(a): kinetic energy (independent of the spin-orbit strength),
(b): spin-orbit energy for $V_0$ = 1000 MeV, and
(c): spin-orbit energy $V_0$ = 2000 MeV.
The lines are the same as in Fig. 2.

The squared overlap between the lowest
$0^+$ state of $^{20}$Ne obtained
by superposing the basis states
with different $R$ and $\Lambda$ values
and each basis state,
which is expressed as
\[
|
\langle
\Psi({\vec r}_1,\cdots,{\vec r}_A;\Lambda, R) 
|
  \Phi({\vec r}_1,\cdots,{\vec r}_A)
\rangle 
|^2
\ / \
\sqrt{ \langle \Psi | \Psi \rangle \langle \Phi | \Phi \rangle  }
\]
in terms of Eqs. (8) and (9),
is shown in Fig. 4
((a): $V_0$ = 0 MeV,
(b): $V_0$ = 1000 MeV, and
(c): $V_0$ = 2000 MeV).
Here, the lines are the same as in Fig. 2.
Both the bra 
(each basis state on the horizontal axis)
and ket (the lowest $0^+$ state 
obtained after superposing the basis states)
states are normalized,
however, since the basis states with
different $\Lambda$ value are 
non-orthogonal, the integration over $\Lambda$
does not correspond to definite physical value.
It is shown that when the spin-orbit interaction
is strong (in the case of (c)),
the peak positions shift to the finite
$\Lambda$ values similarly to Fig. 2.
This result suggests the mixing of two components 
(cluster limit and shell limit) in the ground state
of $^{20}$Ne.
Also, the height of the second peak 
for the solid line ($R = 0.5$ fm)
shown around $\Lambda \sim 2$
increases with increasing strength of the spin-orbit interaction
((a) $\to$ (b) $\to$ (c)).
The states with $\Lambda$ larger than 1 surely correspond to 
particle-hole excitation to higher shells, however 
physical interpretation for it is still an open question.

In Fig. 5, the energy curves of the $0^+$ state 
as a function of the distance between ``quasi cluster"
and $^{16}$O is shown (solid line: total energy, dotted line: spin-orbit energy).
Here, the strength of the spin-orbit interaction
is $V_0$ = 1000 MeV, which gives reasonable level
spacing in Fig. 1. The optimal $\Lambda$ value is chosen 
at each point on the horizontal axis. 
It is shown that the energy becomes minimum around the 
distance of 3 fm, where the contribution of the
spin-orbit interaction is $\sim -2.5$ MeV,
and the expectation value of the spin-orbit interaction
increases to $\sim -8$ MeV in the inner region.

\begin{figure}
\includegraphics[width=4cm]{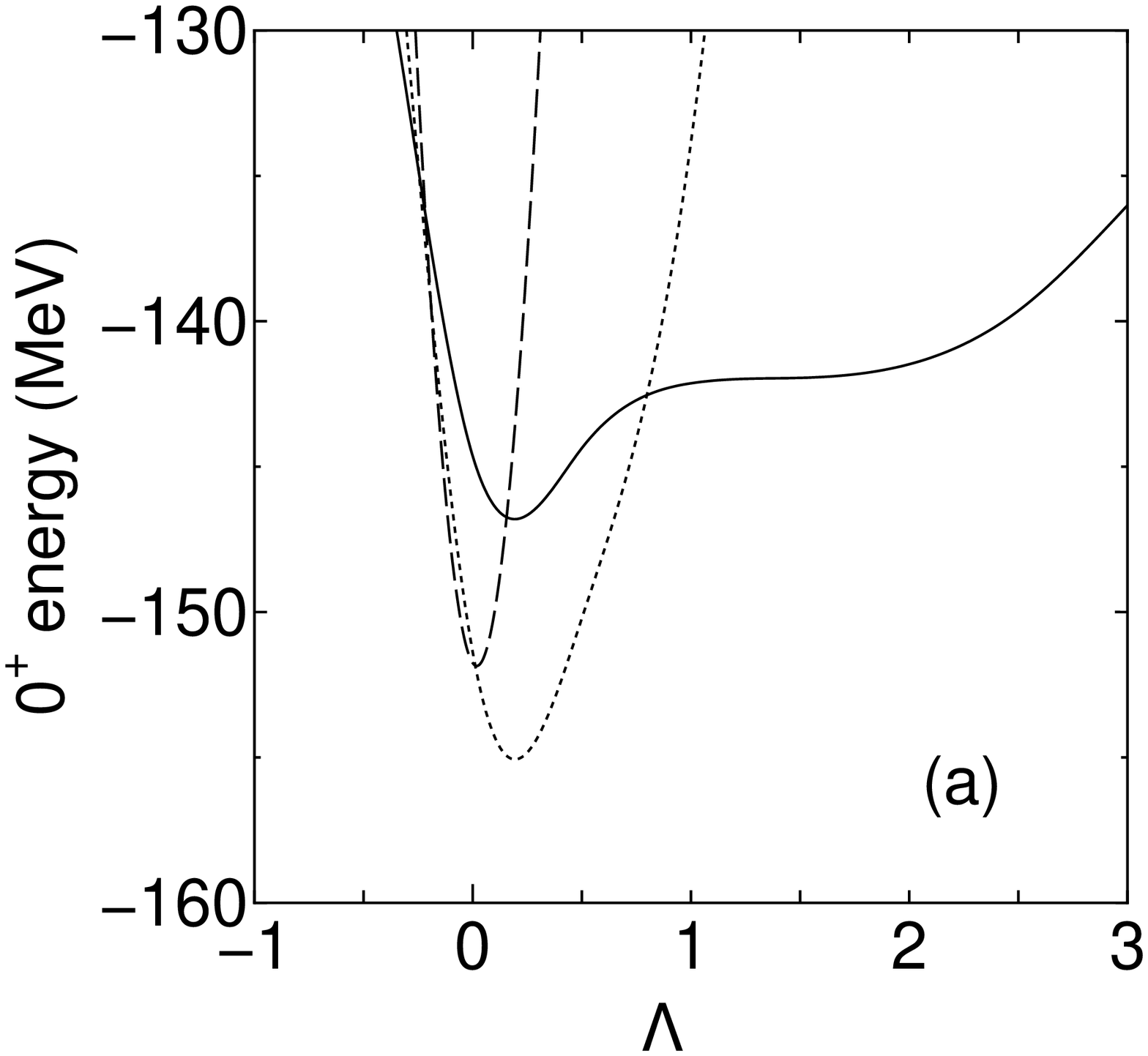} 
\includegraphics[width=4cm]{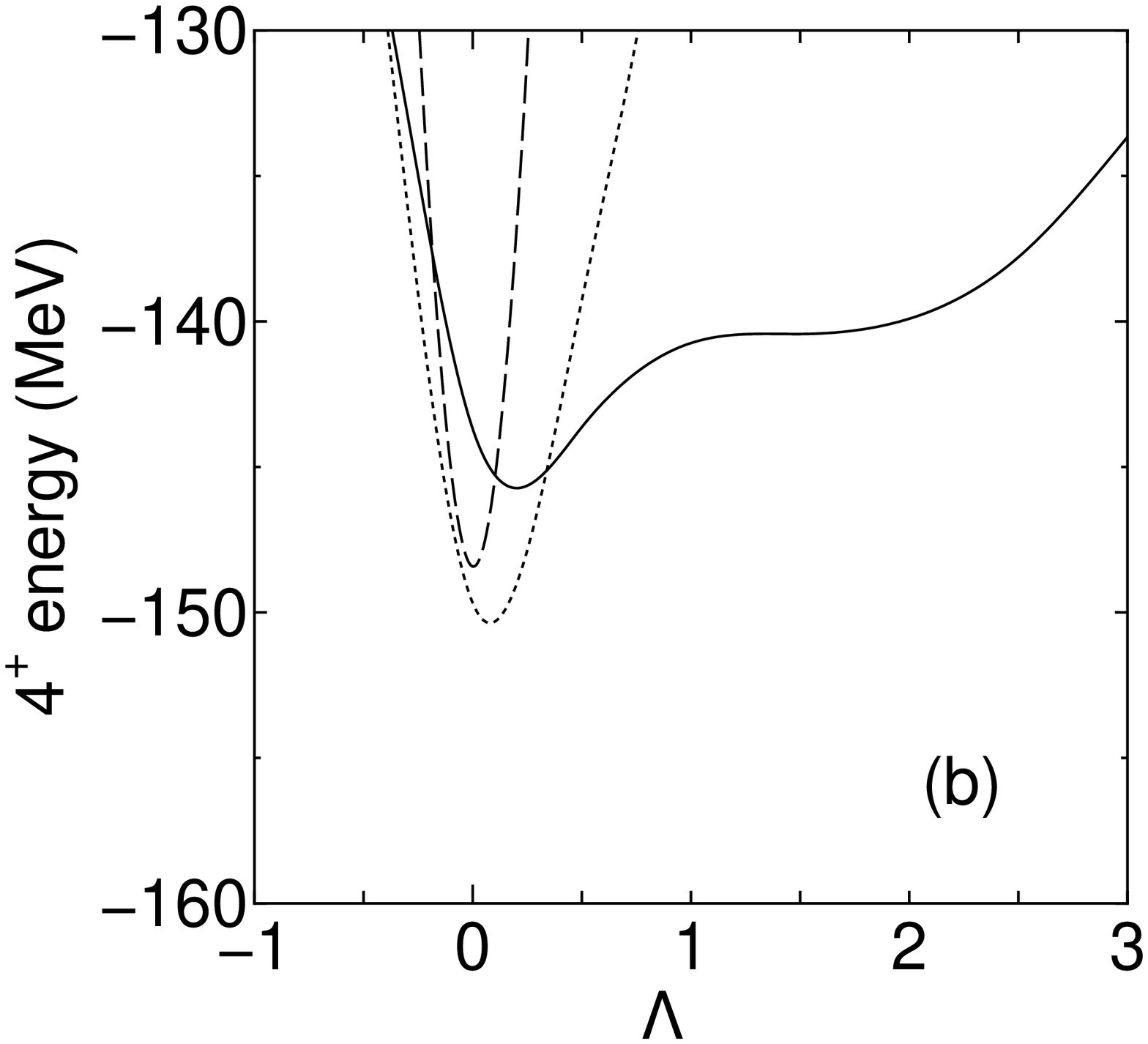}
\includegraphics[width=4cm]{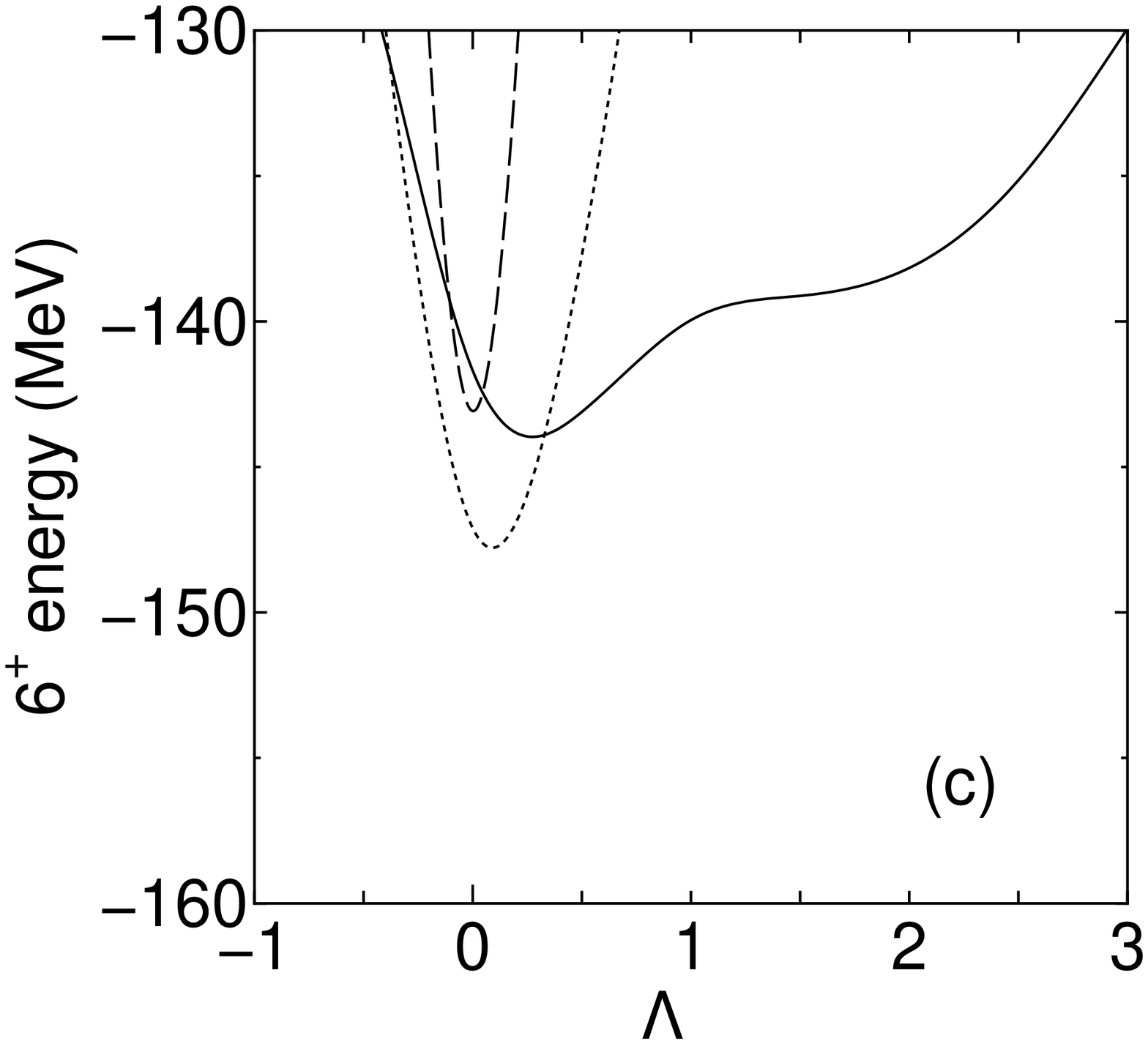}
\includegraphics[width=4cm]{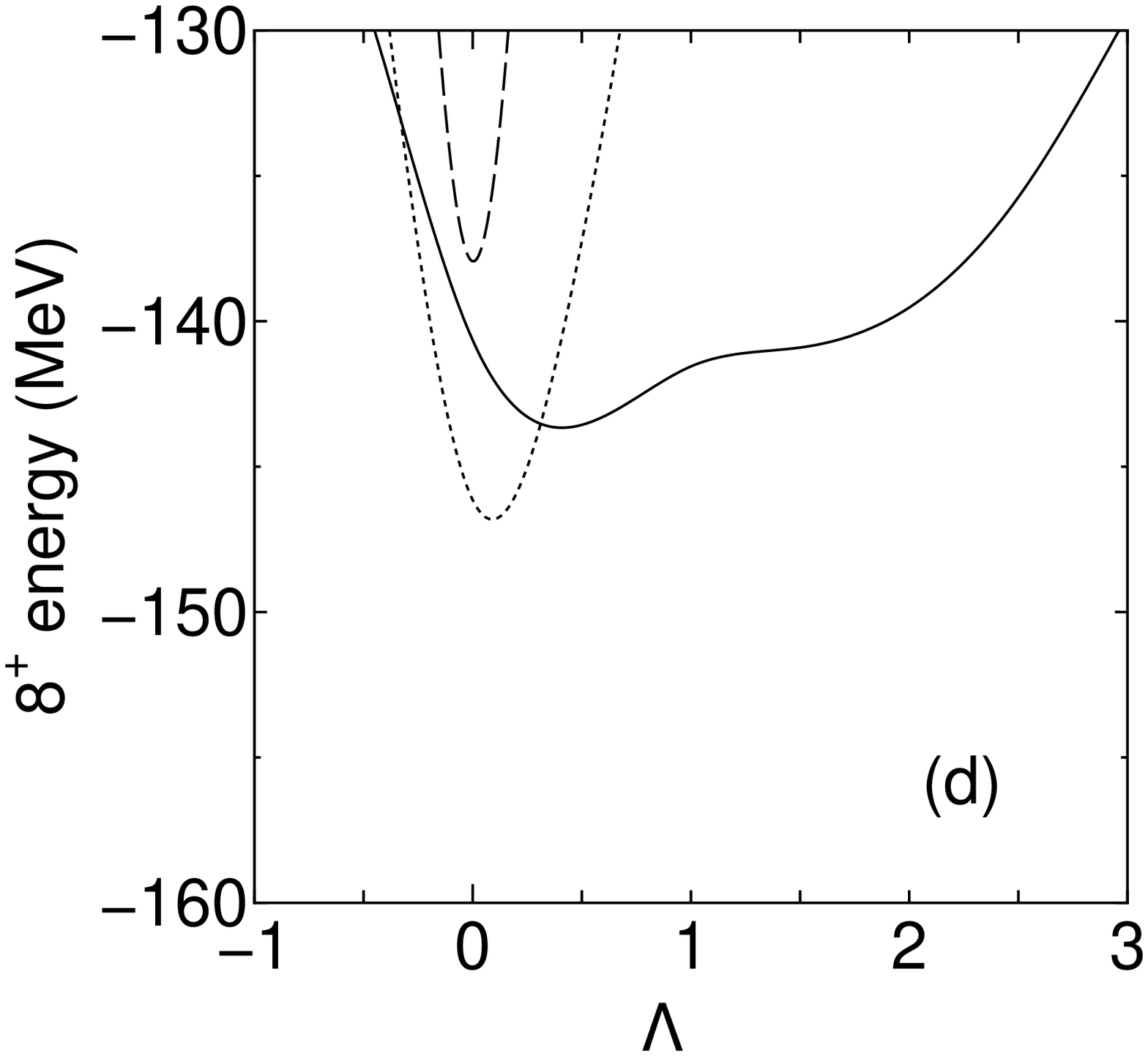}   
\caption{
The energy curves of (a) $2^+$ (b)
$4^+$ (c) $6^+$ and (d) $8^+$ 
states as a function
of $\Lambda$.
The lines are the same as in Figs. 2, 3, 4.
The strength of the spin-orbit interaction is chosen
to be $V_0 = 1000$ MeV.}
\end{figure}

The energy curves of (a) $2^+$, (b)
$4^+$, (c) $6^+$, and (d) $8^+$ 
states as a function
of $\Lambda$ are shown in Fig. 6.
The solid, dotted, 
and dashed lines show the results 
of $R = $ 0.5, 2, and 4 fm, respectively.
Here, the strength of the spin-orbit interaction
is $V_0$ = 1000 MeV, which gives reasonable level
spacing in Fig. 1.
With increasing angular momentum
the optimal $R$ value gets shorter,
and the minimum energy of the dashed line ($R$ = 4 fm)
rapidly increase compared with the solid and dotted lines
in high angular momentum states.
Also, the optimal $\Lambda$ value increases
with increasing angular momentum.

\begin{figure}
\includegraphics[width=4cm]{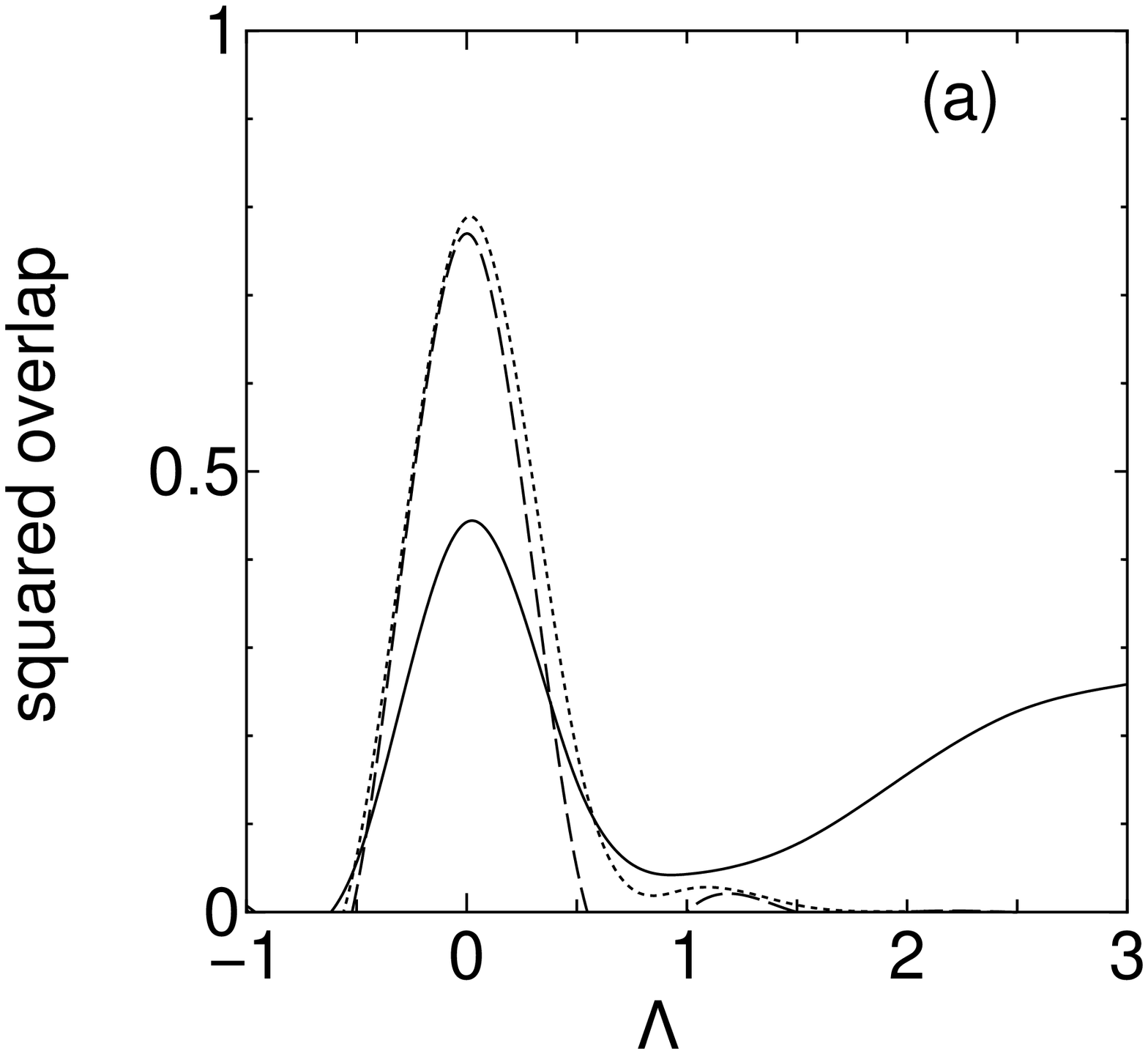} 
\includegraphics[width=4cm]{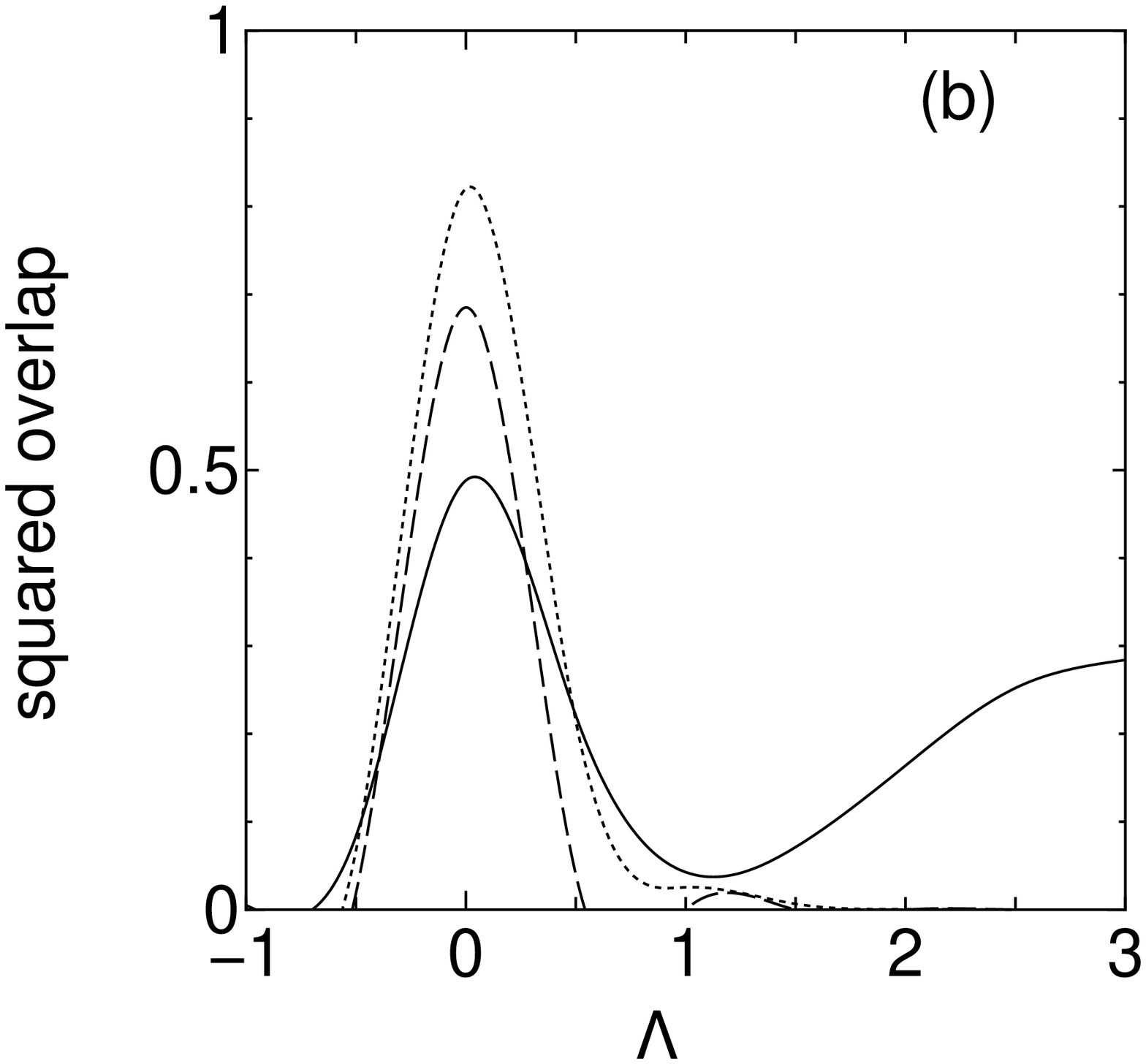}
\includegraphics[width=4cm]{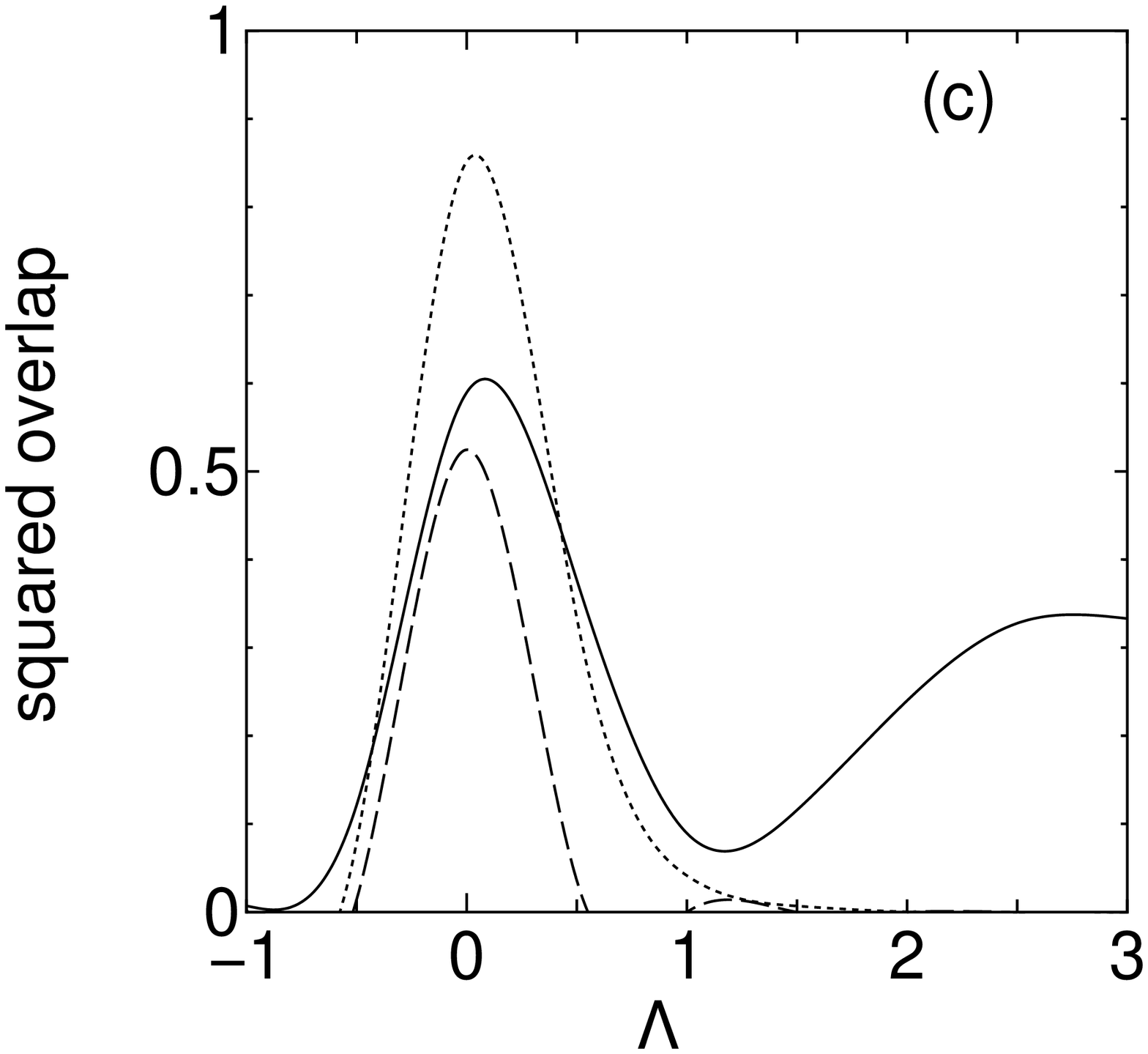}
\includegraphics[width=4cm]{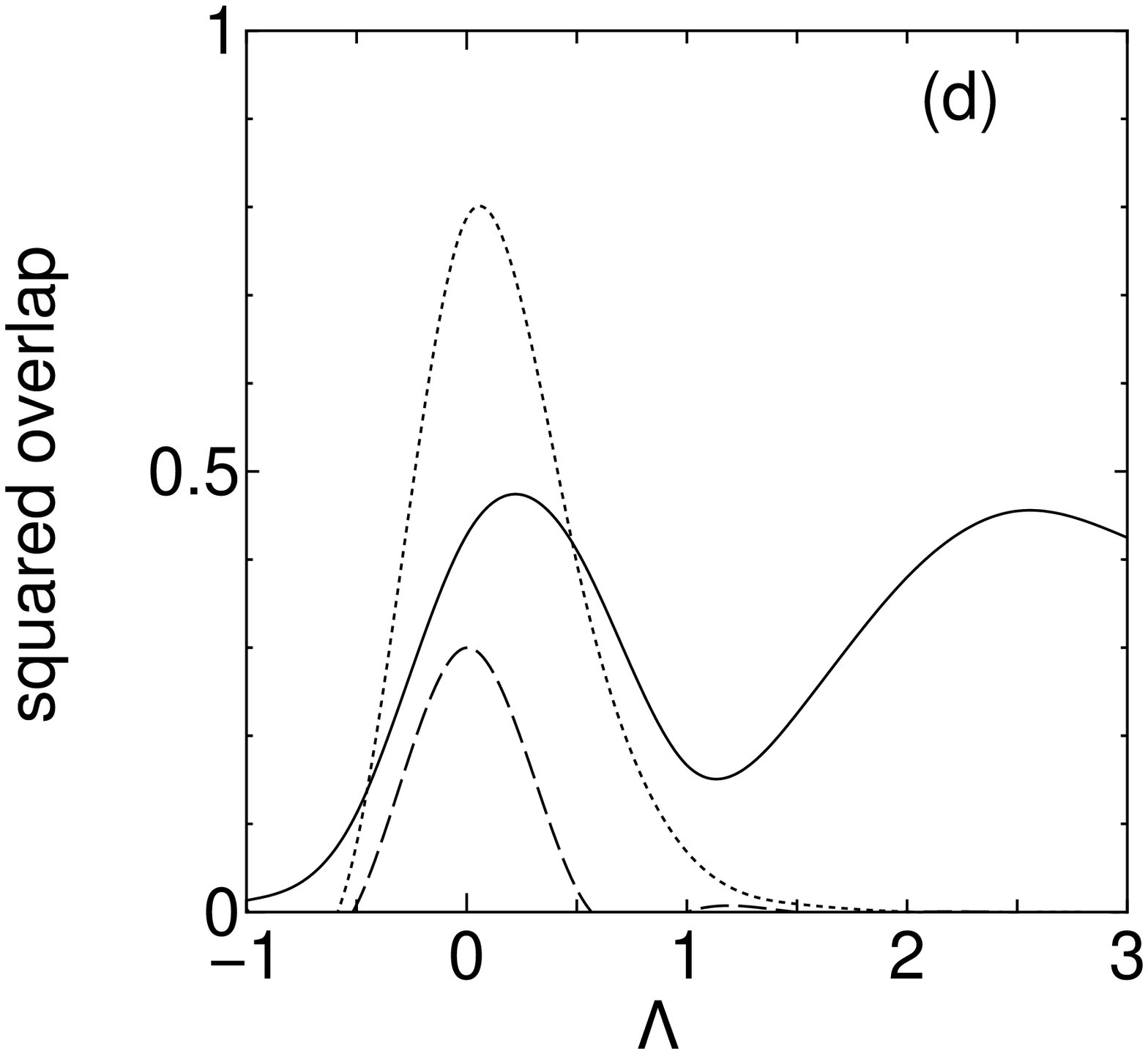}   
\caption{
The squared overlap between the lowest
(a) $2^+$ (b) $4^+$ (c) $6^+$ and (d) $8^+$
states obtained by superposing the basis states
with different $R$ and $\Lambda$ values
and each basis state.
The lines are the same as in Figs. 2, 3, 4, 6.
The strength of the spin-orbit interaction is chosen
to be $V_0 = 1000$ MeV.
}
\end{figure}

The squared overlap between the obtained
(a) $2^+$, (b) $4^+$, (c) $6^+$, and (d) $8^+$
states by superposing the basis states
with different $R$ and $\Lambda$ values
and each basis state is shown in Fig. 7.
The lines are the same as in Figs. 2, 3, 4, 6.
The strength of the spin-orbit interaction is chosen
to be $V_0 = 1000$ MeV.
With increasing angular momentum,
the height of the dashed line ($R$ = 4 fm) 
decreases and the $R$ value which gives 
the largest overlap gets shorter,
similarly to Fig. 6.
Also, the $\Lambda$ value which gives 
the largest overlap slightly increases
with increasing angular momentum.

We can apply the present method also for the negative parity states.
However,  it is known that 
the cluster structure is really important for the lowering of the higher nodal states like the negative parity states. 
Thus the $\alpha$ breaking effect plays a less important role for the negative parity states. 
If we include the $\alpha$ breaking wave functions (finite $\Lambda$ values) in the model space, 
the energy gain for the $1^-$ state of $^{20}$Ne (the band head of the negative parity band) 
is only 0.3 MeV with respect to the pure cluster model configuration, 
for the spin-orbit interaction strength $V_0 $= 1000 MeV. 
This energy gain becomes 0.8 MeV for $V_0$ = 2000 MeV, what is anyway much smaller than in the case 
of the positive parity states.

\subsection{$^{24}$Mg}
Next we discuss the case with two quasi clusters
around the $^{16}$O core, that is $^{24}$Mg.
In the previous case of $^{20}$Ne, one quasi cluster is
placed on the $x$-axis and four nucleons in this
quasi cluster are changed to $Y_{22}$ 
(proportional to $(x+iy)^2/r^2$) or
$Y_{2-2}$ (proportional to $(x-iy)^2/r^2$)
orbitals after giving the imaginary part in the $y$ component. 
In $^{24}$Mg, the second
quasi cluster should be transformed to $Y_{21}$ and $Y_{2-1}$
for spin-up and spin-down nucleons, respectively,
to make the orbital and spin components parallel.
The spherical harmonics
$Y_{21}$ ($Y_{2-1}$) is proportional to $(x+iy)z/r^2$
($(x-iy)z/r^2$), and it is necessary to shift the Gaussian center
parameters of nucleons in quasi clusters also to $z$ direction.

For the first quasi cluster,
we give the Gaussian centers parameter 
\begin{equation}
{\vec \zeta}_1/\sqrt{\nu} = 
R (\vec e_x + i \Lambda \vec e_y + \vec e_z/2)
\end{equation}
for the spin-up proton and neutron, and
\begin{equation}
{\vec \zeta}_1/\sqrt{\nu} = 
R (\vec e_x - i \Lambda \vec e_y + \vec e_z/2)
\end{equation}
for the spin-down proton and neutron.
For the second quasi cluster,
we give
\begin{equation}
{\vec \zeta}_2/\sqrt{\nu} = 
R (\vec e_x + i \Lambda \vec e_y - \vec e_z/2).
\end{equation}
for the spin-up proton and neutron and
\begin{equation}
{\vec \zeta}_2/\sqrt{\nu} = 
R (\vec e_x - i \Lambda \vec e_y - \vec e_z/2)
\end{equation}
for the spin-down proton and neutron,
where $\vec e_x$, $\vec e_y$, and $\vec e_z$ are unit vectors.
The antisymmetrization effect allows us to take 
linear combinations of orbitals for the first 
and the second quasi clusters: when the linear 
combination is in-phase
\begin{equation}
\exp[-\nu(\vec r-\vec \zeta_1/\sqrt{\nu})^2]
+
\exp[-\nu(\vec r-\vec \zeta_2/\sqrt{\nu})^2],
\end{equation}
there is no node in the $z$ direction, however 
when it is anti-phase,
\begin{equation}
(\exp[-\nu(\vec r-\vec \zeta_1/\sqrt{\nu})^2]
-
\exp[-\nu(\vec r-\vec \zeta_2/\sqrt{\nu})^2])/R,
\end{equation}
factor $z$ is multiplied and
one node in the $z$ direction appears at the limit of $R \to 0$.
When we expand the exponents of Eqs. (21) and (22),
similar discussion to Eqs. $(13)-(16)$ follows.
Here the components of $0s$ and $p$ shells disappear
due to the presence of the $^{16}$O core.
Eventually proton and neutron that occupy a state of Eq. (21) 
correspond to  $Y_{22}$ (spin-up case) and $Y_{2-2}$ (spin-down case)
the same as in the $^{20}$Ne case, however proton and neutron
that occupy a state of Eq. (22) correspond to $Y_{21}$ (spin-up case)
and $Y_{2-1}$ (spin-down case) 
due to the factor ``$z$". 
After setting all these Gaussian center parameters,
the whole system is Galilei transformed so that
the center of gravity coincides with the origin of 
the coordinate system.

\begin{figure}
\includegraphics[width=6.5cm]{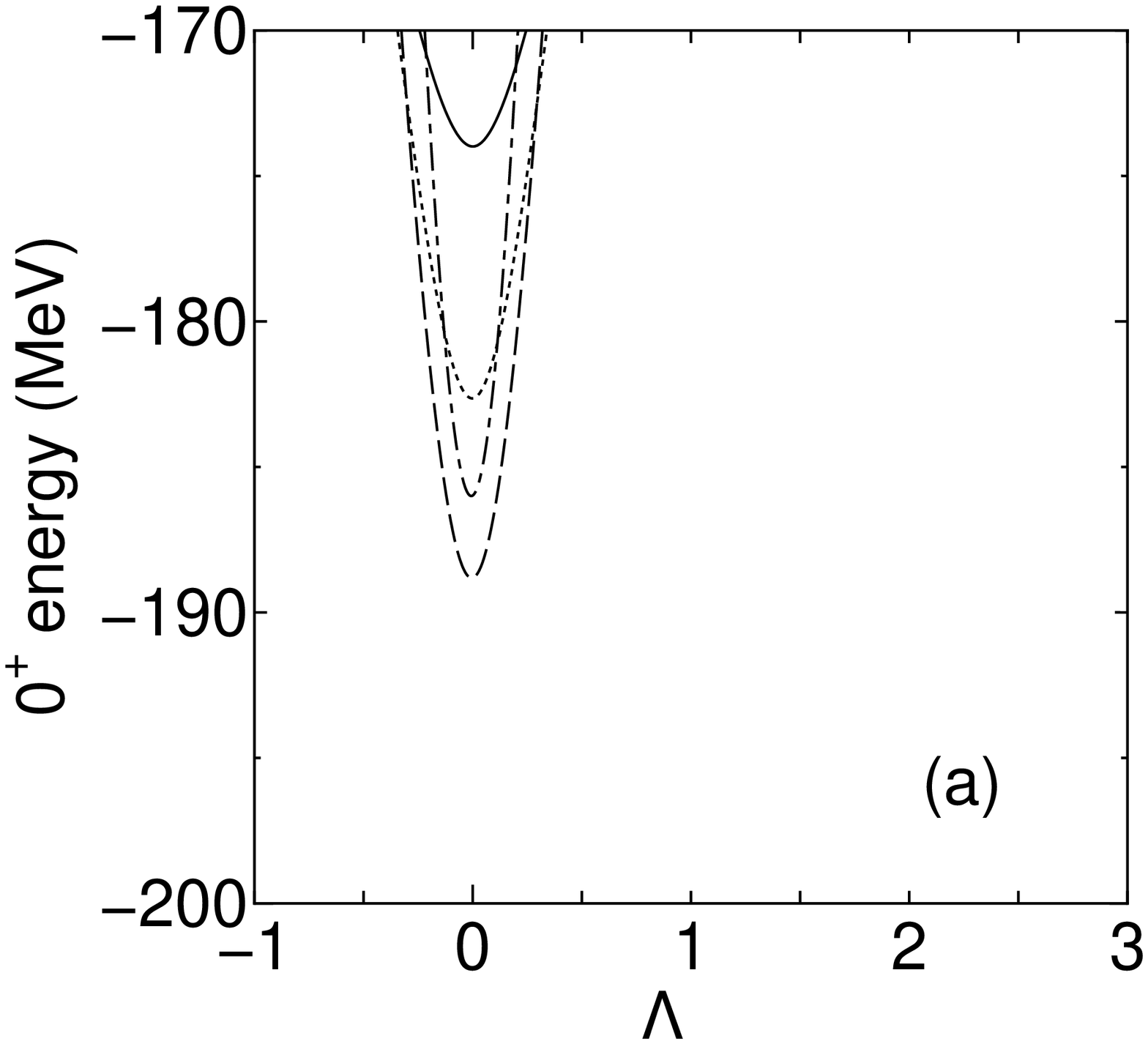} 
\includegraphics[width=6.5cm]{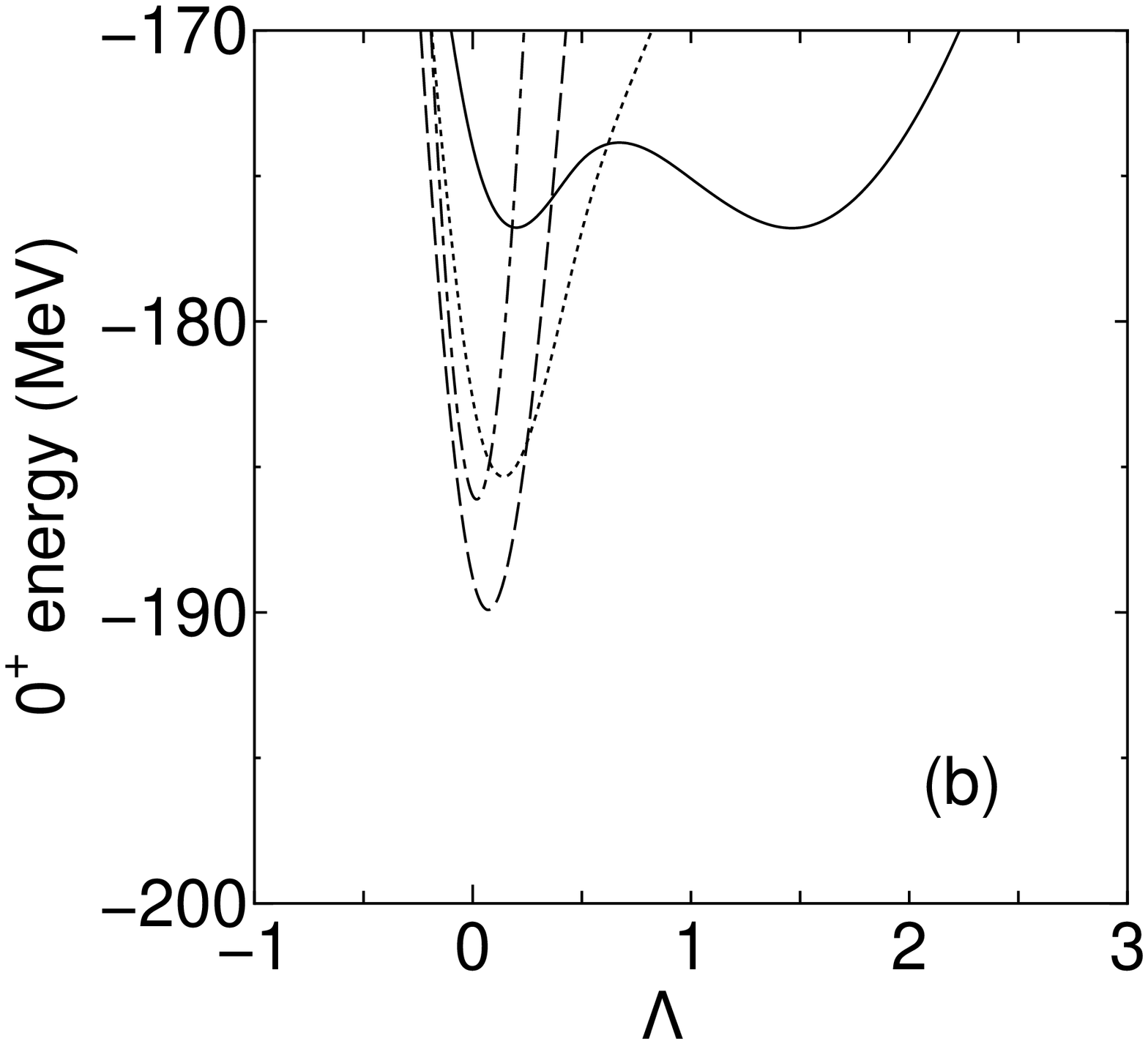}
\includegraphics[width=6.5cm]{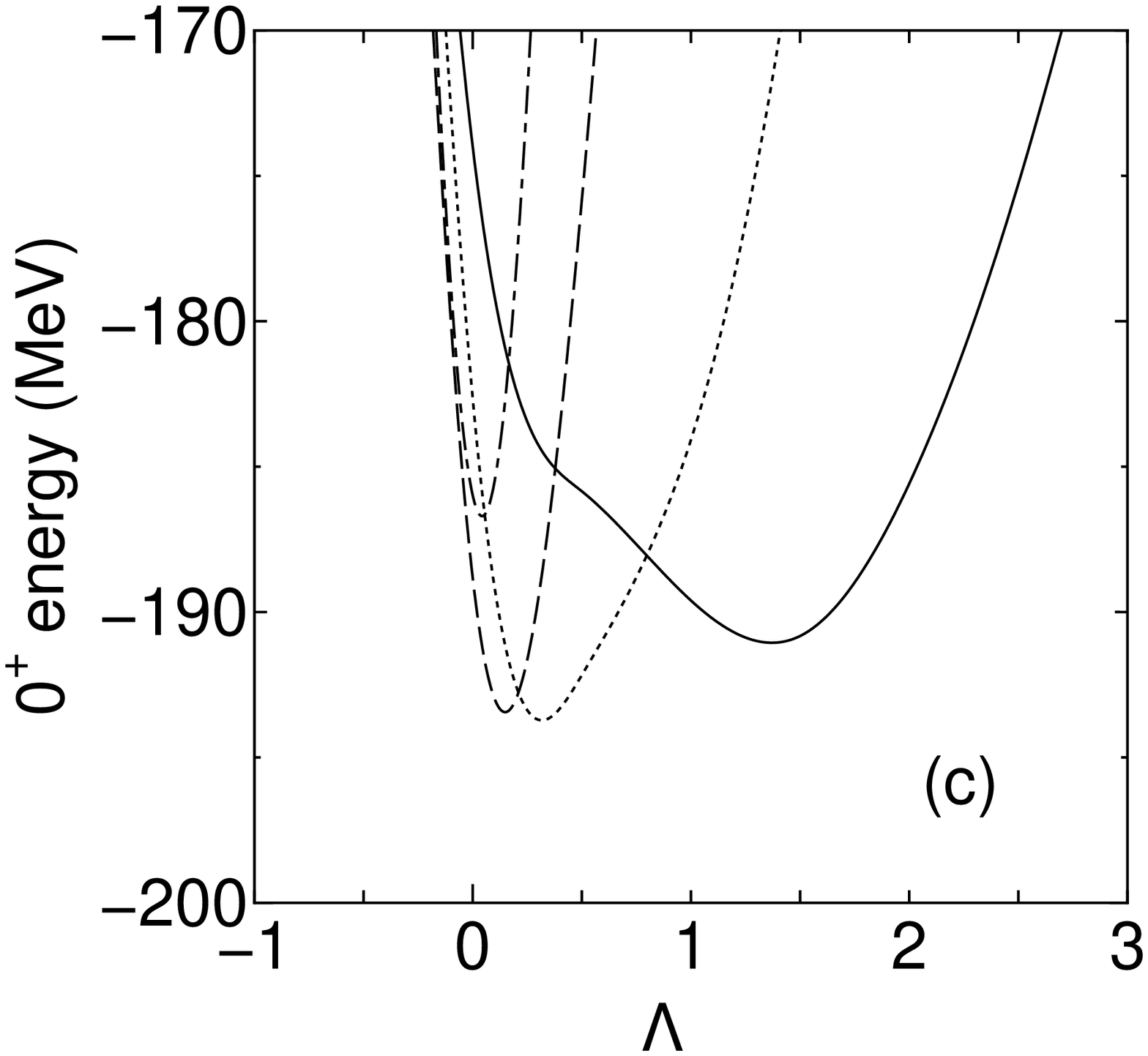}  
\caption{
The energy curves for the
$0^+$ state of $^{24}$Mg as a function of $\Lambda$:
(a): $V_0$ = 0 MeV,
(b): $V_0$ = 1000 MeV, and
(c): $V_0$ = 2000 MeV.
The solid, dotted, dashed, and dash-dotted lines show
$R$ = 0.5, 1, 2, and 3 fm, respectively.}
\end{figure}

The energy curves for the
$0^+$ state of $^{24}$Mg as a function of $\Lambda$
are shown in Fig. 8, where
(a): $V_0$ = 0 MeV,
(b): $V_0$ = 1000 MeV, and
(c): $V_0$ = 2000 MeV.
In Fig. 8 (a), the energy minima are shown around $\Lambda = 0$
independent of $R$, however states with finite $\Lambda$ values
become important especially for lines with small $R$ 
in Figs. 8 (b) and 8 (c) with increasing spin-orbit strength.
In Fig. 8 (b), the second minimum point appears around 
$\Lambda = 1.5$ for the solid line ($R$ = 0.5 fm), and
two local minima of the solid line in Fig. 8 (b) 
merge into a deep minimum point in Fig. 8 (c).

The squared overlap between the lowest
$0^+$ state of $^{24}$Mg obtained
by superposing the basis states
with different $R$ and $\Lambda$ values
and each basis state is shown in Fig. 9
((a): $V_0$ = 0 MeV,
(b): $V_0$ = 1000 MeV, and
(c): $V_0$ = 2000 MeV).
Here, the lines are the same as in Fig. 8.
it is again shown that when the spin-orbit interaction
is strong (in the case of (c)),
the peak positions shift to the finite
$\Lambda$ values similarly to Fig. 4.
This result suggests the mixing of two components 
(cluster limit and shell limit) in the ground state
of $^{24}$Mg.
Also, the height of the second peak 
for the solid line ($R = 0.5$ fm)
shown around $\Lambda \sim 2$
increases with increasing strength of the spin-orbit interaction
((a) $\to$ (b) $\to$ (c)).

\begin{figure}
\includegraphics[width=6.5cm]{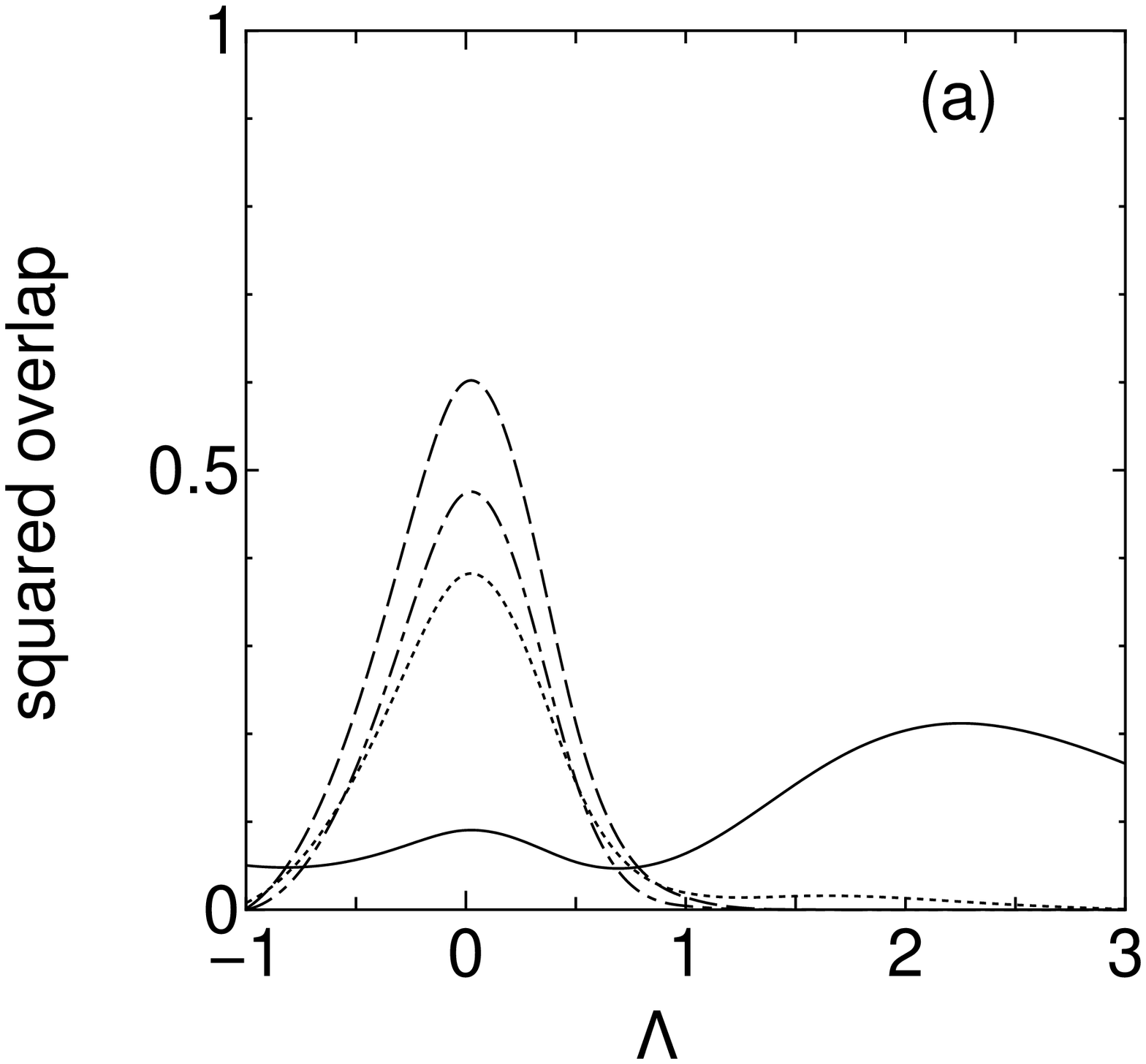} 
\includegraphics[width=6.5cm]{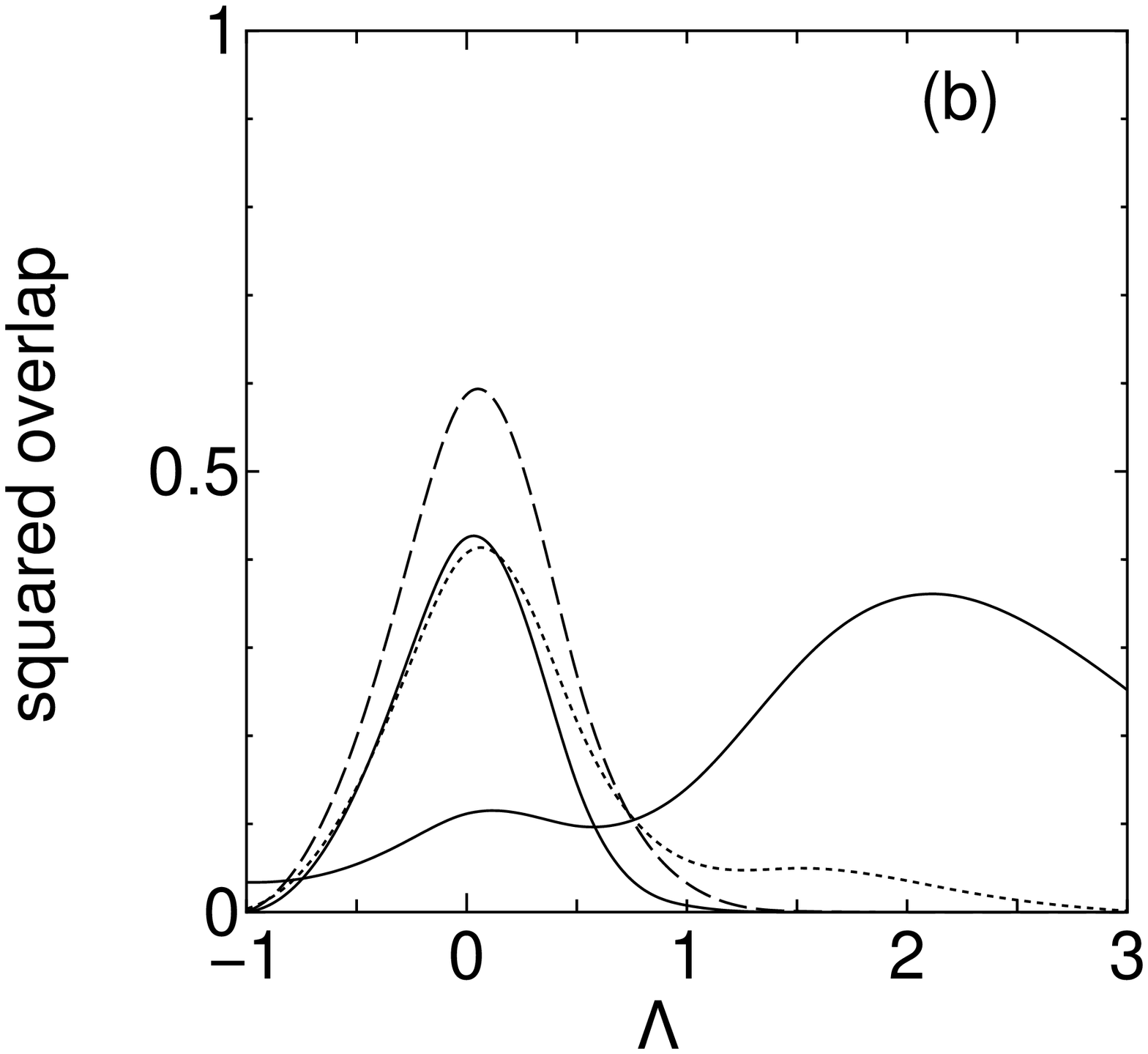}
\includegraphics[width=6.5cm]{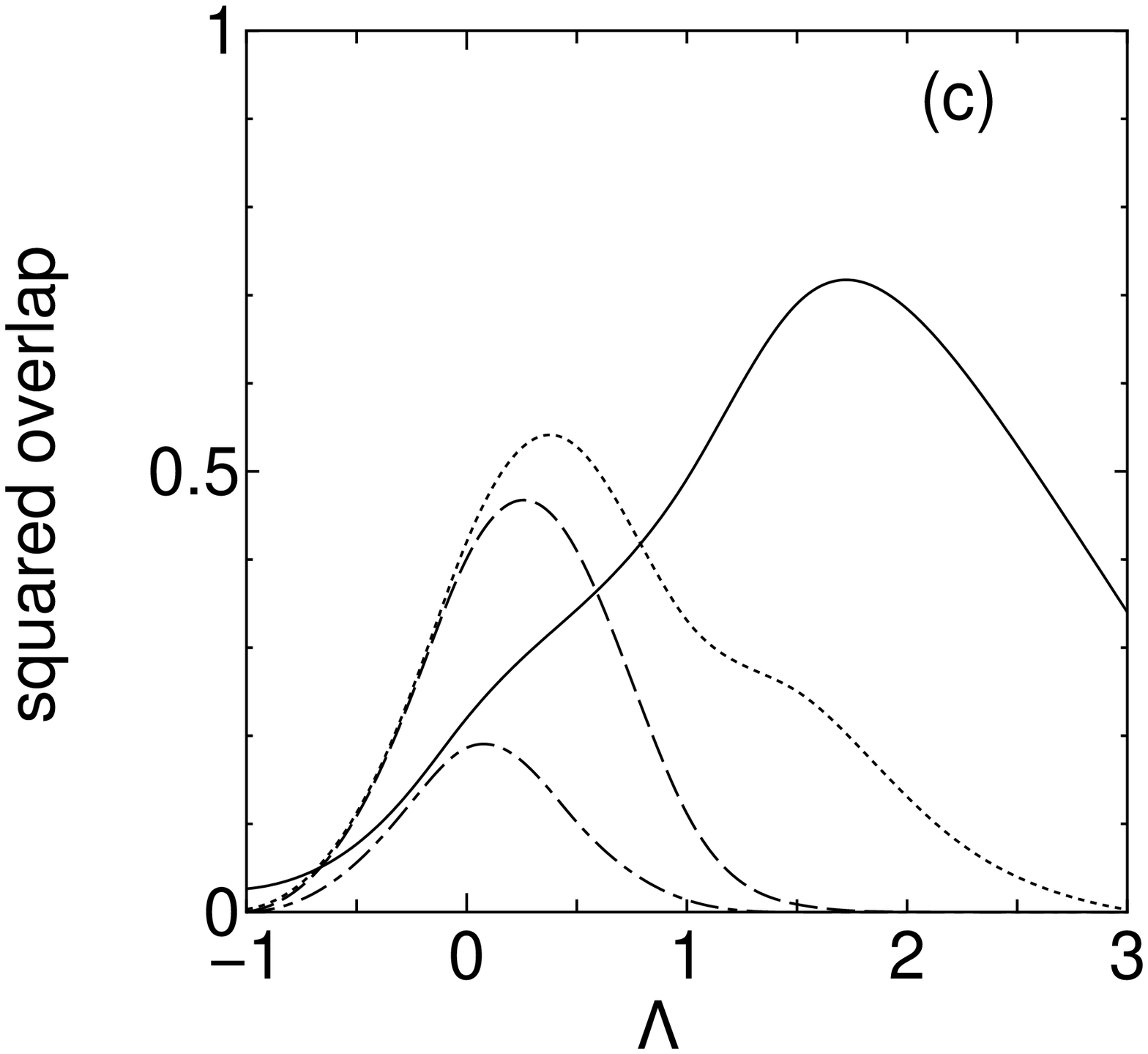}  
\caption{
The squared overlap between the lowest
$0^+$ state of $^{24}$Mg
obtained by superposing the basis states
with different $R$ and $\Lambda$ values
and each basis state.
(a): $V_0$ = 0 MeV,
(b): $V_0$ = 1000 MeV, and
(c): $V_0$ = 2000 MeV.
The lines are the same as in Fig. 8.}
\end{figure}

\begin{table}[h]
\caption{The ground state energy of $^{24}$Mg
calculated with the strength of the spin-orbit interaction,
$V_0$ = 0, 1000, and 2000 MeV. The results
of one and two generator coordinate(s) are compared.
\label{tab:24Mg}}
\begin{tabular}{|r|c|c|}
\hline
$V_0$ & one generator coordinate & two generator coordinates \\
\hline
   0 &  $-189.8$ &  $-190.1$\\ 
1000 & $-190.9$ &  $-191.4$\\ 
2000 & $-196.8$ &  $-198.2$\\ 
\hline
\end{tabular}
\end{table}

For more precise calculation, we can introduce two generator coordinates;
the relative distance between the two quasi clusters ($R_1$ = 0.5, 1.0, 2.0, 3.0 fm) 
and the relative distance between the center of mass of two quasi clusters and
the $^{16}$O core ($R_2$ = 0.5, 1.0, 2.0, 3.0 fm).
The result of the ground state ($0^+$) energy as a function
of the strength of the spin-orbit interaction ($V_0$ in Eq. (7))
is shown in Table I (in the column of two generator coordinates),
together with the results of one generator coordinate ($R$).
Here, the difference between the results of $V_0 = 0$ MeV
and $V_0 = 2000$ MeV is 8.2 MeV in the case of two generator coordinates. 
The effect of the decrease of energy due to the spin-orbit interaction
is doubled compared with $^{20}$Ne (Fig. 1), since we have broken
two $\alpha$ clusters in $^{24}$Mg.

\section {A possible connection to group theoretical approaches}

In this section we discuss how a similar study 
could be formulated in terms of
group theoretical models, in order to investigate the 
possible phase-transitions from a different angle and in a
more quantitative way. 
A reasonable approach would consist of two steps. 
The first one is the algebraic cluster
model calculation, and the second step is the shell-model calculation.
The two parts have a matching point: 
the $U(3)$ dynamical symmetry is present in both models
\cite{varna9}.

\subsection{Basic features of the group theoretical investigations}

The basic features of the 
algebraic studies can be summarized as follows.
A group-theoretical model is considered with a well-defined model space. 
On the other hand the interactions (but only the interactions) are varied. 
They have limiting cases, called dynamical symmetries.
When a dynamical symmetry holds, the eigenvalue-problem has an analytical
solution, due to the fact that the Hamiltonian can be expressed in terms of the
invariant operators of a chain of nested subgroups. In such a case the 
eigenstates of the Hamiltonian have a complete set of good quantum numbers. 
The general Hamiltonian, however, which has
contributions from interactions with different dynamical symmetries, has to be 
diagonalized numerically. The relative weight of the dynamically symmetric 
limits serves as a control parameter, and it defines the 
phase-diagram of the system. When there are more than two dynamical symmetries,
more than one control parameters appear.

In the limit of large particle number 
phase-transitions are seen in the sense that the
derivative of the energy-minimum, as a function of the control-parameter, 
is discontinuous. The order of the derivative, showing the
discontinuity, gives the order of the phase-transition. Thus, in this framework
the phase-transition is investigated quantitatively, like in the
thermodynamics. A phase is defined as a region of the phase diagram between the
endpoint of the dynamical symmetry and the transition point. It is also
conjectured 
\cite{rowephase} 
that such a quantum phase is characterized by a quasi-dynamical
symmetry. Therefore, although the real dynamical symmetry is valid only at a 
single point of the phase-diagram, the more general quasi-dynamical symmetry 
may survive, and in several cases does survive
\cite{rowephase,varna9}, 
in the whole phase. 
If this conjecture really turns out to be true, then 
the situation is similar to Landau's theory: different
phases are determined by different (quasi-dynamical) symmetries, and phase
transitions correspond to a change of the symmetry. 

In the case of the finite particle number the discontinuities are smoothed out, 
as the consequence of the finite size effect, but still remarkable
changes can be detected in the behavior of the corresponding functions.

In this paper we have investigated the cluster-shell competition by enlarging the
cluster-model-space with some shell-model basis states. The shell-model
description, in general, is based on a horizontal truncation of the model space 
(e.g. to a single major shell), while the cluster model makes a vertical 
truncation (e.g. a few nucleons with specific symmetry in many major shells).
Therefore, a model space containing both subspaces, and being a space for
an irreducible representation of a dynamical algebra can be prohibitively large.
Thus, the formulation of the cluster-shell competition in terms of the
group theoretical framework seems to be feasible in a two-step procedure. 
The first step is the cluster study based on an algebraic model, 
the second one is the shell-model calculation. The reason why they can be
combined to a coherent investigation is that they have a matching point: both 
of them have a limit of the $U(3)$ dynamical symmetry
\cite{varna9}.

\subsection{$U(3)$ limit in the algebraic cluster model}

We begin with the algebraic cluster model.
In case of a binary cluster-configuration of closed-shell clusters, 
e.g. $^{16}$O+$^{4}$He,
we practically need to describe only the relative motion.
Algebraically it can be done by applying the vibron model
of $U(4)$ structure
\cite{vib}, 
which have ($l=1$) $\pi$ and and ($l=0$) $\sigma$ bosons,
as building blocks. 
The Pauli-principle can be taken into account by a 
basis-truncation in the simple case of closed-shell clusters
\cite{sacm}, 
and with this constraint the vibron model
can really give a reasonable approach to the $\alpha$-cluster bands 
of $^{20}$Ne
\cite{20neu4}.  

The $U(4)$ algebra of the relative motion
can be considered as the
dynamical algebra of the truncated harmonic oscillator problem
for a finite spectrum (as the compact form of the 
noncompact $U(3,1)$ dynamical algebra of 
the oscillator problem without truncation). 
It is worth stressing here that this simplified algebraic structure 
is valid only from the viewpoint of the physical operators 
when no coupling to the internal cluster degrees of freedom is considered. 
From the viewpoint of the model space, on the other hand,
the internal degrees of freedom have to be taken into account,
otherwise the Pauli-principle could not be appreciated 
\cite{sacm}. 
The internal cluster structure is usually described
in terms of the Elliott-model of $U^{ST}_c(4) \otimes U_c(3)$ group structure,
where $U^{ST}_c(4)$ refers to the spin-isospin degrees of freedom
and $U_c(3)$ stands for the space-part of the internal cluster
wave function.

The cluster model with the $U(4)$ algebraic structure has two dynamical 
symmetries: $U(3)$, and $O(4)$.
Therefore, the phase-space is one-dimensional, having these limits as ending
points. The $U(3)$ limit corresponds to a soft vibrator in the language of
the collective model, or to a shell-model-like clusterization from the
microscopic viewpoint. The $O(4)$ represents the rigid rotor limit, i.e a rigid
molecule-like configuration, which extends to very many major shells \cite{coc8}. 
A recent schematic calculation \cite{hjp}
showed that at a certain value of the control parameter
a phase-transition takes place, and the quasidynamical $U(3)$ symmetry
characterizes the whole phase from the limit of the real dynamical symmetry
up to very close to the point of transition.

When the coupling between the relative motion and internal cluster
degrees of freedom is taken into account 
\cite{sacm},
then  there is a third dynamical symmetry of the cluster
model, denoted by $SO(3)$, which corresponds to the weak-coupling limit.
Then the phase diagram is two dimensional, and can be illustrated by a 
triangle
\cite{varna9}.

\subsection{$U(3)$ limit in the shell model}

The $U(3)$ symmetry is known to be very important in the shell model
since the pioneering work of Elliott
\cite{elli}.
A specific $U(3)$ symmetry defines a (collective) rotational band
in terms of the basis of the spherical shell model.
(In the shell-model context the symmetry is usually called $SU(3)$,
and not $U(3)$. The two groups are uniquely related to each other,
once the total number of particles (oscillator quanta) is given,
this being the only extra generator of $U(3)$ in comparison with $SU(3)$.
In the shell model calculations usually only a single major shell is 
incorporated, therefore, this number does not play any role, and it 
is not needed. In the cluster model, however, several major shells 
are included, thus it is more practical to use $U(3)$, which defines 
$SU(3)$, too.)

The problem of a shell-model-like phase-transition  has been studied in \cite{rbw}. 
This is also a schematic calculation, in which a finite fermion
system was considered, with a Hamiltonian having two dynamical symmetries:
the $SU(3)$ one of the Elliott model, 
and the $SU(2)$ dynamical symmetry, which diagonalizes 
the pairing interaction 
of the superconductivity.
 A phase-transition was observed at a critical point of the
control parameter, and the quasi-dynamical $SU(3)$ symmetry 
turned out to be
valid between the transition point and the endpoint of the $SU(3)$ limit.

From the viewpoint of the phase studies it is remarkable that
recently a triangle-like phase diagram has been proposed also for the
shell model
\cite{frjois},
which, in addition to the $SU(3)$ and $SU(2)$ symmetries
has the independent-particle model as the third corner.

As the realistic calculations are concerned, the method of the
works \cite{vargas}
are especially remarkable from our viewpoint.
There the $^{20}$Ne was described within a shell-model, by considering
4 nucleons in the full sd shell, and applying $SU$(3) basis.
The Hamiltonian includes spin-orbit, quadrupole-quadrupole and pairing 
terms as well.

\subsection{Matching of the two models at the $U(3)$ limit}

The $U(3)$ limit of the two (cluster and shell) models 
coincide with each other, as mentioned
before, and will be discussed more in detail later on. 
The difference in the 
physical content, as indicated here 
(vibrational limit in the cluster model
and rotational limit in the shell model) seems to be a
contradiction, but in fact it is not. 
In the shell model the $U(3)$ dynamical symmetry determines a
rotational band from the basis states of a single major shell. 
On the other hand
the vibrational limit of the cluster model is related to major shell
excitations. If the algebraic ($U(3)$) shell model is extended 
by the inclusion 
of major shell excitations, then it has an algebraic structure 
of $Sp(3,R)$ \cite{sp}. 
In the limit of the large particle number, however, it simplifies to a 
$U$(3) boson model 
\cite{u3},
(or contracted symplectic model
\cite{contsymp},)
which contains both the hydrodynamic model (major shell 
excitations corresponding to giant resonances), and the Elliott $SU$(3) model. 
Similarly, the $U$(3) dynamical symmetry of the cluster model incorporates the 
major shell excitations as well as the $SU$(3) structure within the major shells 
\cite{sacm}. 
(In the language of the liquid drop collective model, Elliott's
$SU(3)$ corresponds to the vorticity of the liquid: the (0,0) $SU(3)$-scalar
represents the irrotational flow, while non-$SU(3)$-scalar corresponds to
non-zero vorticity.)

In this article we have introduced AQC model to describe the cluster-shell competition,
where $\Lambda$ and $R$ are two parameters to characterize the state.
We can map the values of $\Lambda$ and $R$ on the diagram in Fig. 10.
The $U(3)$ limit is characterized as small $R$ and small $\Lambda$,
thus the wave function has both natures of cluster and shell models.
Increasing $\Lambda$ changes $\alpha$ cluster(s) to $jj$-coupling wave
function, which corresponds to the shift of the wave function
from the $U(3)$ limit (common intersection) to the $SU(2)$ limit on the shell model side,
and increasing $R$ corresponds to the change from the $U(3)$ limit
to the $O(4)$ limit (rigid rotor state) on the cluster model side. 
In this way, AQC model can be interpreted in terms of the group theory
by introducing a diagram in Fig. 10.

An algebraic study, which is similar to our present investigation of the
cluster-shell competition would require two kinds of calculations.
The first one is an extension of the work of Ref. \cite{hjp} within 
the algebraic cluster study
for the realistic description of the cluster band(s).
This could determine quantitatively, where the real nucleus sits on the phase
diagram of the cluster model. The $U(3)$ end of this diagram is the common
intersection with the phase diagram of the shell model. In particular, in the
$U(3)$ limit the cluster wave function of the ground state band of $^{20}$Ne e.g.
is given by the shell-model wave function with the quantum numbers:
$U^{ST}(4): [1,1,1,1]$, $S=0$, $T=0$, $U(3): (8,0)$, 
i.e. the shell-model and cluster model wave functions are the same
(apart from the normalization factor).
The second part of the calculation should be a shell-model study,
similar in spirit to those of
\cite{rbw,vargas}.
We discuss here the necessary scenario from the angle of the work
\cite{vargas}.
i) Include not only the ground-band, but also the 
negative parity band, and ii) investigate the behavior of the
system under the influence of the systematic change of the strength
of the spin-orbit and/or pairing interactions. iii) When doing so
the survival of the quasidynamical $SU(3)$ could be determined from
the analysis of the wave function.

The control parameter is the relative weight of the $U(3)$ symmetry-breaking
interaction in both cases. On the cluster side it is the dipole interaction 
(compared to the $U(3)$-symmetric quadrupole one), while on the shell-model 
side it can be spin-orbit and pairing forces. The quasi-dynamical $U(3)$
symmetry defines a phase on both sides of real $U(3)$ dynamical symmetry
(being the matching point of the two phase-diagram). 
This kind of investigation
could answer a question, like: where the system is located on the 
phase diagram of the cluster model and that of the shell-model.
Furthermore, one could figure out if phase-transition(s) take(s)
place, and if so, what kind of transitions.

\begin{figure}
\includegraphics[width=6.5cm]{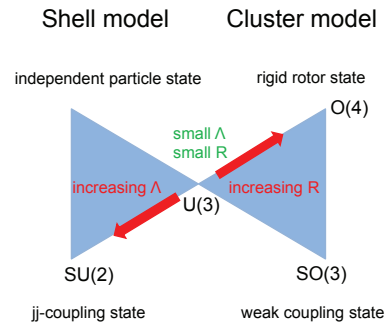}  
\caption{ (Color online)
Schematic diagram for the group theoretical understanding
for cluster-shell competition.
}
\end{figure}

\section{Conclusion and further outlook}

Recently, a microscopic calculation of the cluster-shell
competition became possible, and in this work we introduced the GCM+AQC formalism
to describe this competition in a simple way. The spin-orbit interaction
is a key quantity which drives the transition from CM states to SM states. 
In our model, this effect is implanted into the trial wave function.
By introducing a single parameter $\Lambda$,
one $\alpha$ cluster orbiting around $^{16}$O 
is changed into a ``quasi cluster", where the motion of four nucleons
depends on the spin direction of each nucleon.

The energy levels of $^{20}$Ne calculated in the GCM on a collective basis of the 
AQC states for different strengths of the spin-orbit interaction:
$V_0 = 0, 1000, 2000$ MeV, were compared with the experimental
data. The experimental level spacing of $0^+-2^+-4^+$ shows a clear deviation 
from the $l(l+1)$ rule, and the spin-orbit interaction is needed to explain this deviation.
Although a reasonable description of the scattering phase shift in $^4$He+$N$ reaction
is possible for $V_0 = 2000$ MeV, this value should be reduced for $^{20}$Ne. 
Judging from the level spacing between the ground state $0^+$ and the $8^+$ excited 
state, the optimal value of the spin-orbit coupling strength is $V_0 = 1000$ MeV.
Thus, the spin-orbit strength is overestimated if the scattering phase 
shift in $^4$He+$N$ is fitted without introducing the tensor interaction explicitly.
Both the experimental results and the GCM+AQC calculation
show that the $8^+$ state in $^{20}$Ne deviates strongly from 
the rotational sequence. Our studies demonstrate that the deviation increases 
with increasing the strength of the spin-orbit interaction.

For $V_0 = 0$ MeV, the energy of a $0^+$ state as a function
of $\Lambda$ exhibits a minimum at $\Lambda = 0$. With an increasing value of $V_0$, 
the minimum is shifted towards a finite value of $\Lambda$ and the kinetic energy of 
nucleons in the $\alpha$ quasi cluster increases as well. 
For $V_0 = 1000$ MeV, the energy minimum is found at $\Lambda \sim$ 0.2.   
The optimal value of $\Lambda$ value increases with increasing the angular momentum.

The curves of energy as a function of $\Lambda$ 
show that an optimal value of $R$ (a distance between $^{16}$O
and an  $\alpha$ quasi cluster) is around 3 fm 
for $2^+$. With increasing angular momentum, the optimal value of $R$ decreases
($R$ = 2 fm for $8^+$).

Furthermore, we discussed the case of two quasi clusters
around the $^{16}$O core (the case of $^{24}$Mg) showing the way two 
$\alpha$ clusters can be broken in AQC formalism. 

Concerning the three different kinds of states from the viewpoint of their 
shell or cluster nature: the SM states, the CM states and the SMC states,  the following 
correspondence can be made in the language of the AQC formalism. 
The SM state (type (i)) corresponds to a situation of 
small $R$ and  large $\Lambda$, while a
rigid molecule-like CM state (type (ii)) corresponds to 
large $R$ and small $\Lambda$.
The SMC state (type (iii)) involves 
small $R$ and small $\Lambda$ values. Usually, $R$ is finite due to the
Pauli exclusion principle. 
It follows from the present GCM+AQC study that even parity states of $^{20}$Ne and $^{24}$Mg are 
of the type (iii).

We have also discussed how a similar study could be formulated in terms of
algebraic models.  A possible approach would consist of two steps. 
The first one is the description of experimental data
in terms of the algebraic cluster model, and the second step is the shell-model calculation. 
The reason why they can be combined is that they have a matching point,  
the  limit of the $U(3)$ dynamical symmetry \cite{varna9}, which is the intersection of  
two phase-diagrams. The control parameter would be the relative weight 
of the $U(3)$ symmetry-breaking interactions.  Via this route, one might hope to understand whether 
the phase-transition possibly takes place between the cluster- and 
shell-configurations and what could be the nature of this transitions.

The $U(3)$ symmetry is a common intersection of the cluster and shell models.
CM and SM states can be characterized using the two triangles (see Fig. 10) which merge in the $U(3)$ limit.
In this paper, we have studied the cluster-shell competition using the GCM+AQC microscopic framework.
In this framework, $\Lambda$ and $R$ are parameters of a collective AQC basis, which
give an insight into the nature of many-body states. These parameters can be associated with features shown 
on this diagram. The $U(3)$ limit is characterized by small $R$ and small $\Lambda$ values,
thus the wave function has the nature of both cluster and shell models.
Increasing $\Lambda$ changes $\alpha$ cluster(s) into $jj$-coupling wave
function, which corresponds to the shift from $U(3)$ to the $SU(2)$ limit on the shell model side.
Finally, increasing $R$ corresponds to the change from $U(3)$ to the $O(4)$ limit 
(the rigid rotor state) on the cluster model side. In this way, the microscopic results of the GCM+AQC model 
can be interpreted in intuitive language of the group theory.

\begin{acknowledgments}
One of the authors (N.I.) would like to thank
JSPS Core-to-core program. This work was supported in part by the OTKA
grant K72357, MNiSW grant N N202 033837, and COPIGAL.
\end{acknowledgments}


\begin{references}
\bibitem{Boh75} 
A. Bohr and B.R. Mottelson, {\em Nuclear Structure} (Benjamin, New York, 1975), Vol II.
\bibitem{Jensen} 
O. Haxel, J.H.D. Jensen, and H.E. Suess, Phys. Rev. {\bf 75}, 1766 (1949).
\bibitem{Mayer} 
M.G. Mayer, Phys. Rev. {\bf 75}, 1969 (1949).
\bibitem{Ike68}
K. Ikeda, N. Takigawa, and H. Horiuchi, Prog. Theor. Phys. Suppl. Extra number, 464 (1968).
\bibitem{Oko03}
J. Oko{\l}owicz, M. P{\l}oszajczak, and I. Rotter, Phys. Rep. {\bf 374}, 271 (2003);\\
R. Chatterjee, J. Oko{\l}owicz, and M. P{\l}oszajczak, Nucl. Phys. A {\bf 764}, 528 (2006);\\
J. Oko{\l}owicz, M. P{\l}oszajczak, and Y. Luo, Acta Phys. Pol. {\bf 39}, 389 (2008);\\
Y. Luo, J. Oko{\l}owicz, and M. P{\l}oszajczak, arXiv:nucl-th/0211068.
\bibitem{Dob07}
J. Dobaczewski, N. Michel, W. Nazarewicz, M. P{\l}oszajczak, and J. Rotureau, Prog. in Part. and Nucl. Phys. {\bf 59}, 432 (2007);\\
J. Oko{\l}owicz, M. P{\l}oszajczak, and Yan-an Luo, Acta Phys. Pol. {\bf 39}, 389 (2008).
\bibitem{Danos} 
M. Danos and B.M. Spicer, Z. Physik {\bf 237}, 320 (1970);\\
A. Arima, V. Gillet, and J. Ginocchio, Phys. Rev. Lett. {\bf 25}, 1043 (1970);\\
M. Danos and V. Gillet, Phys. Lett. B {\bf 34}, 24 (1971);\\
F. Catara, A. Insolia, and U. Lombardo, Nucl. Phys. A {\bf 261}, 282 (1976).
\bibitem{Wheeler} 
J.A. Wheeler, Phys. Rev. {\bf 52}, 1083 (1937);\\
C.F. von Weizs\"acker, Naturwiss. {\bf 26}, 209 (1938);\\
W. Wefelmeier, Z. Phys. {\bf 107}, 332 (1937).
\bibitem{Hiu63}
J. Hiura and I. Shimodaya, Prog. Theor. Phys. {\bf 30}, 585 (1963); {\it ibid.} {\bf 36}, 977 (1966).
\bibitem{Tam65} R. Tamagaki and H. Tanaka, Prog. Theor. Phys. {\bf 34}, 191 (1965).
\bibitem{Brink}
D.M. Brink, {\it In Proceedings of the International School of Physics ``Enrico Fermi" Course XXXVI},
edited by C. Bloch (Academic, New York, 1966), p. 247.
\bibitem{Supple}
Y. Fujiwara, H. Horiuchi, K. Ikeda, M. Kamimura, K. Kat\=o, Y. Suzuki, and E. Uegaki,
Prog. Theor. Phys. Suppl. {\bf 68}, 60 (1980).
\bibitem{Oko} S. Dro\.zd\.z, J. Oko{\l}owicz, and M. P{\l}oszajczak, Phys. Lett. B {\bf 128}, 5 (1983.)
\bibitem{Bau} W. Bauhoff, E. Caurier, B. Grammaticos, and M. P{\l}oszajczak, Phys. Rev. C {\bf 32}, 1915 (1985).
\bibitem{FMD}
T. Neff, and H. Feldmeier, Nucl. Phys. A {\bf 738}, 357 (2004).
\bibitem{CC} 
For example, {\it Proc. of the 8th Int. Conf. on Clustering Aspects of Nuclear Structure and Dynamics}, Edited by K. Ikeda, I. Tanihata, and H. Horiuchi, Nucl. Phys.  A {\bf 738} (2004).
\bibitem{Oertzen} 
W. von Oertzen, Z. Phys. A {\bf 354}, 37 (1996); A {\bf 357}, 355 (1997).

\bibitem{Freer} 
M. Freer $et\ al.$, Phys. Rev. Lett. {\bf 82}, 1383 (1999). 
\bibitem{vOe06}
W. von Oertzen, M. Freer, and Y. Kanada-En'yo, Phys. Rep. {\bf 432}, 43 (2006).
\bibitem{Vol06} A. Volya and V. Zelevinsky, Phys. Rev. C{\bf 74}, 064314 (2006).            
\bibitem{Mic09} 
N. Michel, W. Nazarewicz, M. P{\l}oszajczak, and T. Vertse, J. Phys. G: Nucl. Part. Phys. {\bf 36}, 013101 (2009).
\bibitem{coc8} J. Cseh, J. Darai, A. Algora, H. Yepez-Martinez, and P. O. Hess, 
            Rev. Mex. Fis. S {\bf 54} (3), 30 (2008). 
\bibitem{CS}
N. Itagaki, S. Aoyama, S. Okabe, and K. Ikeda, Phys. Rev. C {\bf 70}, 054307 (2004).
\bibitem{SSS}
N. Itagaki, A. Kobayakawa, and S. Aoyama, Phys. Rev. C {\bf 68}, 054302 (2003).
\bibitem{Gri57} J.J. Griffin and J.A. Wheeler, Phys. Rev. {\bf 108}, 311 (1957).
\bibitem{Simple}
N. Itagaki, H. Masui, M. Ito, and S. Aoyama, Phys. Rev. C {\bf 71} 064307 (2005). 
\bibitem{Masui}
H. Masui, N. Itagaki, Phys. Rev. C {\bf 75} 054309 (2007). 
\bibitem{phase-rev} 
P. Cejnar and F. Iachello, J. Phys. A {\bf 40}, 581 (2007);\\
J. Cseh, J. Darai, H. Yepez-Martinez, and P. O. Hess, Int. J. Mod. Phys. E {\bf 17}, 2296 (2008);\\
P. Cejnar and J. Jolie, Progr. Part. Nucl. Phys. {\bf 62}, 210 (2009);\\
F. Iachello, AIP Conf. Proc. {\bf 1165}, 193 (2009). 
\bibitem{varna9} J. Cseh, 
J. Phys. Conf. Ser. {\bf 205}, 012021 (2010).
\bibitem{explanation}
These components are necessary because the spin-orbit interaction has the form of $\vec l \cdot \vec s$, and both spatial part ($\vec l$) and spin part ($\vec s$) are time odd.
\bibitem{Ono1}
A. Ono, H. Horiuchi, T. Maruyama, and A. Ohnishi, Prog. Theor. Phys. {\bf 87}, 1185 (1992);
Phys. Rev. Lett. {\bf 68}, 2898 (1992).
\bibitem{FMD1}
H. Feldmeier, Nucl. Phys. A {\bf 515}, 147 (1990).
\bibitem{Drozdz}
S. Dro\.zd\.z, J. Oko{\l}owicz, and M. P{\l}oszajczak, Phys. Lett. B {\bf 109}, 145 (1982);\\
E. Caurier, B. Grammaticos, and T. Sami, Phys. Lett. B {\bf 109}, 150 (1982).
\bibitem{explanation1}
The cluster width $\nu$ becomes a complex parameter and is allowed to vary in the TDC \cite{Drozdz} and FMD \cite{FMD1} approaches. 
\bibitem{FMD2}
H. Feldmeier, K. Bieler, and J. Schnack, Nucl. Phys. A {\bf 586}, 492 (1995).
\bibitem{Koonin}
J. Da Provid\~{e}ncia, Nucl. Phys. {\bf 46}, 401 (1963);\\
D.K. Kerman, and S.E. Koonin, Ann. Phys. {\bf 100}, 332 (1976).
\bibitem{Vol}
A.B. Volkov, Nucl. Phys. {\bf 74}, 33 (1965).
\bibitem{G3RS} 
R. Tamagaki, Prog. Theor. Phys. {\bf 39}, 91 (1968).
\bibitem{Okabe79}
S. Okabe and Y. Abe, Prog. Theor. Phys. {\bf 61}, 1049 (1979). 
\bibitem{Matsuse}
T. Matsuse, M. Kamimura, and Y. Fukushima, Prog. Theor. Phys. {\bf 53}, 706 (1975);\\
Y. Fukushima, M. Kamimura, and T. Matsuse, Prog. Theor. Phys. {\bf 55}, 1310 (1976).
\bibitem{Kato} 
K. Kat\=o and H. Band\=o, Prog. Theor. Phys. {\bf 59}, 774 (1978).
\bibitem{Fujiwara} 
Y. Fujiwara, H. Horiuchi, and R. Tamagaki, Prog. Theor. Phys. {\bf 61}, 1629 (1979);\\
Y. Fujiwara, Prog. Theor. Phys. {\bf 62}, 122 (1979).
\bibitem{Akiyama} 
Y. Akiyama, A. Arima, and T. Sebe, Nucl. Phys. A {\bf 138}, 273 (1969).
\bibitem{McGrory} 
J.B McGrory and B.H. Wildenthal, Phys. Rev. C {\bf 7}, 974 (1973).
\bibitem{Tomoda}
T. Tomoda, A. Arima, Nucl. Phys. A {\bf 303}, 217 (1978).
\bibitem{Suzuki}
S. Hara, K. T. Hecht, and Y. Suzuki, Prog. Theor. Phys. {\bf 84}, 254 (1990);\\
S. Hara, K. Ogawa, and Y. Suzuki,  Prog. Theor. Phys. {\bf 84}, 254 (1992).
\bibitem{EnyoNe}
Y. Kanada-En'yo and  H. Horiuchi, Prog. Theor. Phys. {\bf 93}, 115 (1995).
\bibitem{Taniguchi} Y. Taniguchi, M. Kimura, and H. Horiuchi, Prog. Theor. Phys. {\bf 112}, 475 (2004).




\bibitem{Yamada}
T. Yamada, Phys. Rev. C {\bf 42}, 1432 (1990).
\bibitem{Terasawa}
T. Terasawa, Prog. Theor. Phys. {\bf 23}, 87 (1960); \\
A. Arima and T. Terasawa, Prog. Theor. Phys. {\bf 23}, 115 (1960).
\bibitem{Myo}
T. Myo, K. Kat\=o, and K. Ikeda, Prog. Theor. Phys. 
{\bf 113}, 763 (2005).
\bibitem{Tensor}
T. Otsuka, T. Suzuki, R. Fujimoto, H. Grawe, and Y. Akaishi, 
Phys. Rev. Lett. {\bf 95}, 232502 (2005).
\bibitem{rowephase} D.J. Rowe, Nucl. Phys.  A{\bf 745}, 47 (2004).
\bibitem{vib} F. Iachello, Phys. Rev. C {\bf 23}, 2778 (1981); \\
          F. Iachello and R.D. Levine, J. Chem. Phys. {\bf 77}, 3046 (1982).
\bibitem{sacm} J. Cseh, Phys. Lett. B {\bf 281}, 173 (1991); \\
            J. Cseh and G. L\'evai, Ann. Phys. (NY) {\bf 230}, 165 (1994). 
\bibitem{20neu4} J. Cseh, J. Phys. Soc. Jpn. {\bf 58} Suppl. 604 (1989). 

\bibitem{hjp} H. Yepez-Martinez, J. Cseh, and P. O. Hess, 
           Phys. Rev. C {\bf 74}, 024319 (2006).
\bibitem{elli} J.P. Elliott, Proc. Roy. Soc. {\bf 245}, 128 (1958).
\bibitem{rbw} D.J. Rowe, C. Bahri, and W. Wijesundera, 
          Phys. Rev. Lett. {\bf 80}, 4394 (1998).
\bibitem{frjois} A. Frank, J. Jolie, P. Van Isacker,
          Symmetries in Atomic Nuclei, Springer, New York (2009).
\bibitem{vargas} C. Vargas, J.G. Hirsch, P.O. Hess, J.P. Draayer,
             Phys. Rev. C{\bf 58}, 1488 (1998);\\
             C. Vargas, J.G. Hirsch, J.P. Draayer,
             Nucl. Phys. A{\bf 690}, 409 (2001).
\bibitem{sp} D.J. Rowe, Rep. Prog. Phys.  {\bf 48}, 1419 (1985), 
         and references therein.
\bibitem{u3} G. Rosensteel and D.J. Rowe, Phys. Rev. Lett. {\bf 47}, 223 (1981).
\bibitem{contsymp} O. Castanos, J.P. Draayer, Nucl. Phys. A {\bf 491}, 349 (1989).
\end{references}
\end{document}